\documentclass[5p,times]{elsarticle}
\usepackage{graphicx}
\usepackage{xcolor}
\usepackage{adjustbox}
\usepackage[ruled]{algorithm2e}

\usepackage{amsfonts,amssymb,exscale} 
\usepackage{amsmath}
\usepackage{mathtools}
\usepackage{bm,upgreek}
\usepackage{multirow}
\usepackage[english]{babel}
\usepackage{psfrag}
\usepackage{hyphenat}
\usepackage{mathptmx}   
\usepackage{xcolor}
\usepackage{orcidlink}
\usepackage{multicol}
\usepackage{empheq}
\usepackage{natbib}
\usepackage{array}
\newcolumntype{t}{>{\tt}c}
\newcolumntype{u}{>{\tt}l}
\newcommand{\orcid}[1]{\href{https://orcid.org/#1}{\textcolor[HTML]{A6CE39}{\aiOrcid}}}
\newcommand{\beq}{\begin{equation}}
\newcommand{\eeq}{\end{equation}}
\newcommand{\D}  {\displaystyle}

\def\scas  #1{{\rm{#1}}{}}
\def\vecs  #1{{\rm{\bf{#1}}}{}}
\let\vec\bm
\let\ten\bm
\journal{arXiv}
\begin{document}
\begin{frontmatter}
\title{A universal material model subroutine for soft matter systems}   
\author[lab1]{Mathias Peirlinck\orcidlink{0000-0002-4948-5585}}
\author[lab2]{Juan A. Hurtado\orcidlink{0000-0002-5616-8158}}
\author[lab3]{Manuel K. Rausch\orcidlink{0000-0003-1337-6472}}
\author[lab4]{Adrian Buganza Tepole\orcidlink{0000-0001-8531-0603}}
\author[lab5]{Ellen Kuhl\orcidlink{0000-0002-6283-935X}}
\address[lab1]{Department of BioMechanical Engineering $\cdot$ Faculty of Mechanical Engineering $\cdot$ Delft University of Technology $\cdot$ Delft $\cdot$ the Netherlands}
\address[lab2]{Dassault Systèmes $\cdot$ Providence, RI $\cdot$ USA}
\address[lab3]{Department of Mechanical Engineering $\cdot$ University of Texas at Austin $\cdot$ Austin, TX $\cdot$ USA}
\address[lab4]{Department of Mechanical Engineering $\cdot$ Purdue University $\cdot$ West Lafayette, IN $\cdot$ USA}
\address[lab5]{Department of Mechanical Engineering $\cdot$ Stanford University $\cdot$ Stanford, CA $\cdot$ USA}
\date{April 19th, 2024}
\begin{abstract}
Soft materials play an integral part in many aspects of modern life including autonomy, sustainability, and human health, and their accurate modeling is critical to understand their unique properties and functions.
Today's finite element analysis packages come with a set of pre-programmed material models, which may exhibit restricted validity in capturing the intricate mechanical behavior of these materials.
Regrettably, incorporating a modified or novel material model in a finite element analysis package requires non-trivial in-depth knowledge of tensor algebra, continuum mechanics, and computer programming, making it a complex task that is prone to human error.
Here we design a universal material subroutine, which automates the integration of novel constitutive models of varying complexity in non-linear finite element packages, with no additional analytical derivations and algorithmic implementations.
We demonstrate the versatility of our approach to seamlessly integrate innovative constituent models from the material point to the structural level through a variety of soft matter case studies: a frontal impact to the brain; reconstructive surgery of the scalp; diastolic loading of arteries and the human heart; and the dynamic closing of the tricuspid valve.
Our universal material subroutine empowers all users, not solely experts, to conduct reliable engineering analysis of soft matter systems. 
We envision that this framework will become an indispensable instrument for continued innovation and discovery within the soft matter community at large.
\end{abstract}
\end{frontmatter}
\section{Motivation}
Understanding the mechanical behavior of soft matter is pivotal across various scientific and engineering domains, ranging from biophysics, over soft robotics, to biomedical and material science engineering. 
Biological materials, composites, polymers, foams, and gels all
exhibit complex non-linear mechanical behaviors and functions, which result from the intrinsic architecture and interactions of their constituent molecules or particles. 
To characterize this behavior, a multitude of constitutive material models have been proposed in the literature \cite{He2022}. \\[6.pt]
\noindent Finite element analysis provides a versatile and powerful framework to evaluate these highly nonlinear material models and predict their mechanical response within complex geometries and under various loading conditions.
Most contemporary finite element software packages offer an extensive number of standard isotropic and anisotropic hyperelastic material models, including neo-Hooke \cite{Treloar1948}, Mooney Rivlin \cite{Mooney1940,Rivlin1948}, Ogden \cite{Ogden1972}, or Yeoh \cite{Yeoh1993}.
However, the implementation of newly discovered constitutive models requires the definition of novel material model subroutines or plugins, which map the computational domain's second-order kinematic deformation gradient tensor to a second-order Cauchy stress tensor \cite{Abaqus2024}.
These material subroutines are evaluated within every finite element, at each integration point, within every time step, at each Newton iteration. \\[6.pt]
\noindent Unfortunately, the efficient integration of novel constitutive models into non-linear finite element software packages is a complex task \cite{Kiran2014,Connolly2019}.
The user needs to derive and implement explicit forms of the second-order Cauchy stress tensor and the fourth-order spatial elasticity tensor \cite{Maas2018}.
The derivation and coding of these complicated tensorial expressions can be an extremely hard task \cite{Miehe1996}, and requires a non-trivial deep understanding of tensor algebra, continuum mechanics, computational algorithms, data structures, and software architecture \cite{Fehervary2020}.
Non-surprisingly, such endeavors are highly subject to human errors \cite{Young2010}.
This high degree of effort and risk of human error when integrating novel constitutive models in finite element packages limits its use to expert specialists, and, as such, hampers research progress, dissemination, and sharing of models and results amongst a broad and inclusive community.\\[6.pt]
In this work, we streamline the implementation of novel constitutive models into existing finite element analysis software, and mitigate the risk for human error. We provide a common language and framework for the computational mechanics community at large.
We design a modular and universal material subroutine, which automates the incorporation of constitutive models of varying complexity in non-linear finite element analysis packages and requires no additional analytical derivations and algorithmic implementations by the user.
First, we introduce the concept of constitutive neural networks, which form the architectural backbone for our universal material model.
Next, we illustrate the universal material model itself, 
describe its internal structure through pseudocodes, 
and showcase how this subroutine can be effortlessly integrated and activated within finite element simulations.
We provide specific examples on how existing constitutive models fit in our overarching framework, and how we can incorporate special constitutive cases that feature mixed invariant features.
Finally, we showcase the flexibility of our approach to naturally integrate novel constitutive models from the material point level to the structural level through various soft matter modeling case studies: the mechanical simulation of a frontal impact to the brain, reconstructive surgery of the scalp, the diastolic loading of arteries and the human heart, and the dynamic closing of the tricuspid valve.\\
\label{motivation}
\section{Constitutive modeling}
\label{CNN}
\begin{figure*}[h]
\centering
\includegraphics[width=0.6\linewidth]{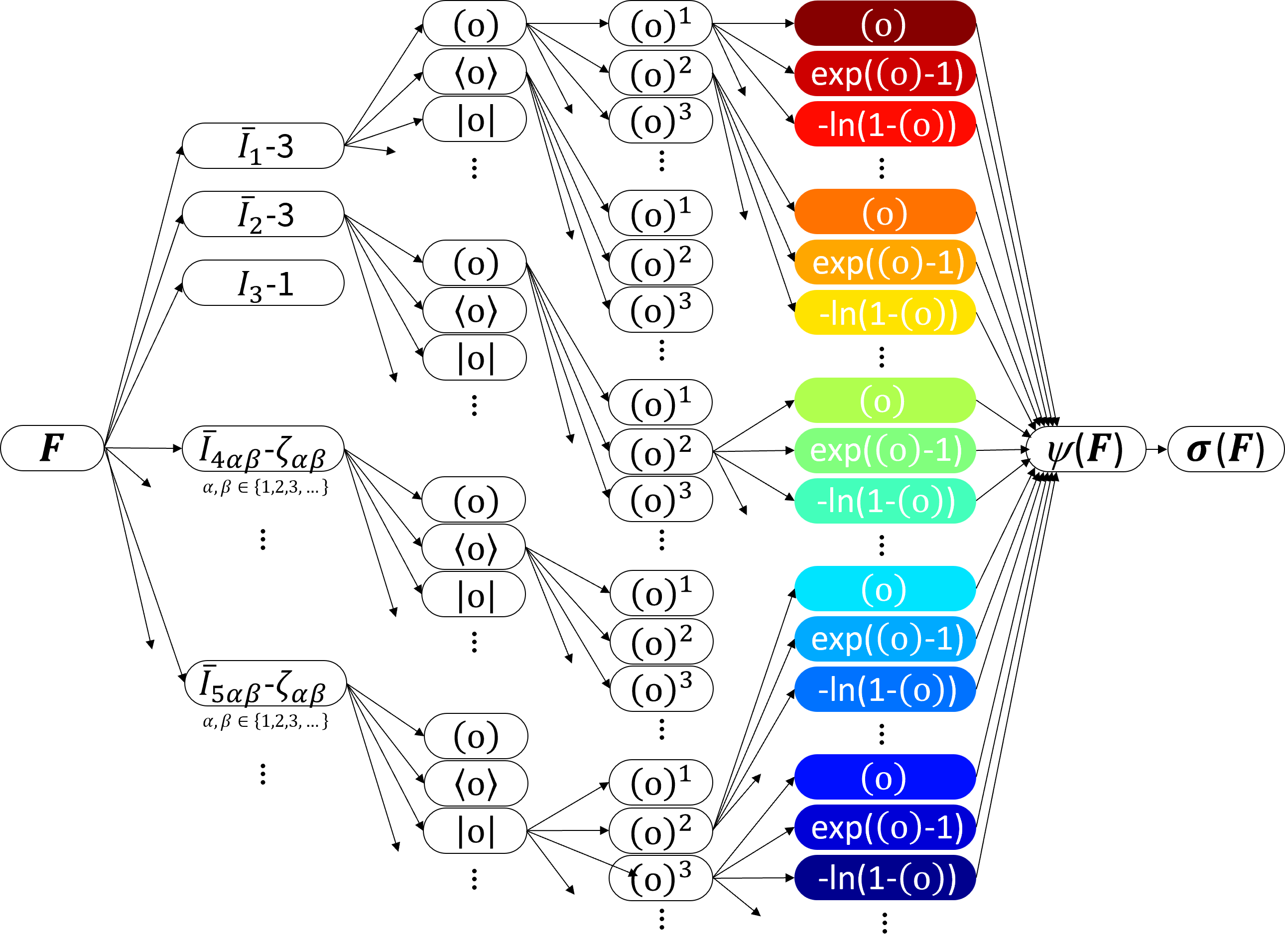}
\caption{{\textbf{\sffamily{Constitutive neural network architecture.}} Anisotropic, compressible, feed forward constitutive neural network with three hidden layers to approximate the single scalar-valued free energy $\psi(\bar{I}_1, \bar{I}_2, I_3, \bar{I}_{\rm{4\alpha\beta}},\bar{I}_{\rm{5\alpha\beta}})$,
as a function of 15 invariants of the left Cauchy-Green deformation tensor $\ten b$. The zeroth layer generates identity $(\circ)$, the rectified linear unit $\langle \circ \rangle$, and the absolute value $\langle \circ \rangle$ of the 15 invariants. The first layer generates powers $(\circ)$, $(\circ)^2$, $(\circ)^3$, etc. and the second layer applies the identity $(\circ)$, the exponential  $(\rm{exp}(\circ)-1)$, and the logarithm $(-\rm{ln}(1-(\circ)))$ to these powers. The network is not fully connected by design to satisfy the condition of polyconvexity \textit{a priori}.
}}\label{figcnnarchitecture}
\end{figure*}
\subsection{Kinematics}
We introduce the deformation map $\vec{\varphi}$ as the mapping of material points $\vec{X}$ in the undeformed configuration to points $\vec{x}=\vec{\varphi}(\vec{X})$ in the deformed configuration \cite{Antman2005,Holzapfel2000book}. The gradient of the deformation map $\vec{\varphi}$ with respect to the undeformed coordinates $\vec{X}$ defines the deformation gradient $\ten{F}$ with its determinant $J$, 
\beq
\ten{F} 
= \nabla_{\vecs{X}} \vec{\varphi}
\quad \mbox{with} \quad
J = \det (\ten{F}) > 0,
\eeq
We multiplicatively decompose the deformation gradient $\ten{F}$ into its volumetric $\ten{F}^{\rm{vol}}$ and isochoric $\bar{\ten{F}}$ parts \cite{Flory1961},
\beq
\ten{F} 
= \ten{F}^{\rm{vol}} \cdot \bar{\ten{F}}
\quad \mbox{with} \quad
\ten{F}^{\rm{vol}} = J^{\frac{1}{3}}\ten{I}
\quad \mbox{and} \quad
\bar{\ten{F}} = J^{-\frac{1}{3}}\ten{F}.
\label{eq:volsplit}
\eeq
As deformation measures, we introduce
the left and right Cauchy-Green deformation tensors, 
$\ten{b}$ and $\ten{C}$,
and their isochoric counterparts,
$\bar{\ten{b}}$ and $\bar{\ten{C}}$,
\beq
\begin{array}{l@{}c@{}l@{}c@{}l}
    \ten{b}
&=& \ten{F}^{\scas{t}} 
&\cdot& \ten{F} \\
    \ten{C}
&=& \ten{F}
&\cdot& \ten{F}^{\scas{t}}
\end{array} 
\qquad \mbox{and} \qquad
\begin{array}{l@{}c@{}l@{}c@{}l}
    \bar{\ten{b}} 
&=& \bar{\ten{F}}^{\scas{t}} 
&\cdot &\bar{\ten{F}}\\
    \bar{\ten{C}} 
&=&\bar{\ten{F}} 
&\cdot &\bar{\ten{F}}^{\scas{t}}\,.
\end{array}
\label{eq:CGdeftensors}
\eeq
We further assume directionally-dependent behavior, with three preferred directions, 
$\vec{n}_{1}^0$, $\vec{n}_{2}^0$, $\vec{n}_{3}^0$, 
associated with the material's internal fiber directions 
in the reference configuration,
where all three vectors are unit vectors,
$||\,\vec{n}_{1}^0\,|| = 1$, 
$||\,\vec{n}_{2}^0\,|| = 1$, 
$||\,\vec{n}_{3}^0\,|| = 1$.
Based on the volumetric and isochoric decomposition, 
and the underlying fiber orientations in the material,
we characterize the deformation in terms of 15 invariants \cite{Spencer1971,Menzel2004}.
More specifically, we define one isotropic volumetric invariant,
\beq
I_3 = \det (\ten{F}^{\scas{t}} \cdot \ten{F}) = J^2 \,,
\label{eq:inv3F}
\eeq
two isotropic deviatoric invariants,
\beq
\begin{aligned}
    &\bar{I}_1 = [\bar{\ten{F}}^{\scas{t}} \cdot \bar{\ten{F}} ] : \ten{I} \\
    &\bar{I}_2 = \mbox{$\frac{1}{2}$} \; [ \bar{I}_1^2 - [\bar{\ten{F}}^{\scas{t}} \cdot \bar{\ten{F}} ] : [\bar{\ten{F}}^{\scas{t}} \cdot \bar{\ten{F}} ] ],\\
\end{aligned}
\label{eq:invisoF}
\eeq
six anisotropic deviatoric invariants,
\beq
\begin{array}{l@{\hspace*{0.4cm}}l}
\bar{I}_{\rm{4(11)}} = [\bar{\ten{F}}^{\scas{t}} \cdot \bar{\ten{F}} ] : [\vec{n}_{1}^0 \otimes \vec{n}_{1}^0]
& \bar{I}_{\rm{5(11)}} = [\bar{\ten{F}}^{\scas{t}} \cdot \bar{\ten{F}} ]^2 : [\vec{n}_{1}^0 \otimes \vec{n}_{1}^0] \\
\bar{I}_{\rm{4(22)}} = [\bar{\ten{F}}^{\scas{t}} \cdot \bar{\ten{F}} ] : [\vec{n}_{2}^0 \otimes \vec{n}_{2}^0]
& \bar{I}_{\rm{5(22)}} = [\bar{\ten{F}}^{\scas{t}} \cdot \bar{\ten{F}} ]^2 : [\vec{n}_{2}^0 \otimes \vec{n}_{2}^0] \\
\bar{I}_{\rm{4(33)}} = [\bar{\ten{F}}^{\scas{t}} \cdot \bar{\ten{F}} ] : [\vec{n}_{3}^0 \otimes \vec{n}_{3}^0]
& \bar{I}_{\rm{5(33)}} = [\bar{\ten{F}}^{\scas{t}} \cdot \bar{\ten{F}} ]^2 : [\vec{n}_{3}^0 \otimes \vec{n}_{3}^0] 
\end{array}
\label{eq:invanisoF}
\eeq
and six deviatoric coupling invariants,
\beq
\begin{array}{l@{\hspace*{0.4cm}}l}
\bar{I}_{\rm{4(12)}} = [\bar{\ten{F}}^{\scas{t}} \cdot \bar{\ten{F}} ] : [\vec{n}_{1}^0 \otimes \vec{n}_{2}^0]
& \bar{I}_{\rm{5(12)}} = [\bar{\ten{F}}^{\scas{t}} \cdot \bar{\ten{F}} ]^2 : [\vec{n}_{1}^0 \otimes \vec{n}_{2}^0] \\
\bar{I}_{\rm{4(13)}} = [\bar{\ten{F}}^{\scas{t}} \cdot \bar{\ten{F}} ] : [\vec{n}_{1}^0 \otimes \vec{n}_{3}^0]
& \bar{I}_{\rm{5(13)}} = [\bar{\ten{F}}^{\scas{t}} \cdot \bar{\ten{F}} ]^2 : [\vec{n}_{1}^0 \otimes \vec{n}_{3}^0] \\
\bar{I}_{\rm{4(23)}} = [\bar{\ten{F}}^{\scas{t}} \cdot \bar{\ten{F}} ] : [\vec{n}_{2}^0 \otimes \vec{n}_{3}^0]
& \bar{I}_{\rm{5(23)}} = [\bar{\ten{F}}^{\scas{t}} \cdot \bar{\ten{F}} ]^2 : [\vec{n}_{2}^0 \otimes \vec{n}_{3}^0] 
\end{array}
\label{eq:invanisocoupleF}
\eeq
Note that these coupling invariants reverse their sign if one of the fiber directions changes its sign, and can therefore not be considered strictly invariant. Nevertheless, these pseudo-invariants were found to be convenient for the definition of anisotropic constitutive models \cite{Holzapfel2009}.
\subsection{Free energy function}
To ensure thermodynamic consistency, we introduce the Helmholtz free energy $\psi$ as a function of the deformation gradient $\psi=\psi \left( \ten{F} \right)$.
Assuming no dissipative energy losses within the material, and rewriting the Clausius–Duhem entropy inequality \cite{Planck1897} following the Coleman and Noll principle \cite{Coleman1959,Gasser2021book}, we derive 
\beq
\ten{\sigma}
=\frac{1}{J}
\frac{\partial \psi\left( \ten{F} \right)}{\partial \ten{F}} \cdot \ten{F}^\scas{t}
\label{eq:cauchyderivF}
\eeq
as the constitutive relation between Cauchy stress $\ten{\sigma}$ and deformation gradient $\ten{F}$.
To guarantee that our free energy function $\psi$ satisfies \textit{material objectivity} and \textit{material symmetry}, we further constrain our stress responses to be functions of the invariants of the left and right Cauchy Green deformation tensors $\ten{b}$ and $\ten{C}$ \cite{Spencer1971,Spencer1984}.
This results in the general definition of the free energy function $\psi$ as a function of the 15 invariants,
\beq
\psi\left( \ten{F} \right) \doteq \psi \left( \bar{I}_1 \, , \bar{I}_2 \, , I_3 \, , \bar{I}_{\rm{4(\alpha\beta)}} \, , \bar{I}_{\rm{5(\alpha\beta)}} \right),
\label{eq:psiinvgeneral}
\eeq
with $\alpha \, , \beta \in \{ 1,2,3 \}$ and $\beta \geq \alpha$.
To account for the quasi-incompressible behavior of soft materials, we make the constitutive choice to additively decompose our free energy function $\psi$ into volumetric $\psi^{\rm{vol}}$ and isochoric $\bar{\psi}$ parts,
\beq
\psi \doteq \psi^{\rm{vol}} + \bar{\psi}.
\label{eq:psitotal}
\eeq
Here, we define the volumetric free energy contribution,
\beq
\psi^{\rm{vol}} = \psi_3(I_3) \,,
\eeq
in terms of the isotropic volumetric invariant $I_3$ (Eq. \ref{eq:inv3F}), 
and the deviatoric free energy contribution,
\beq
\begin{aligned}
\bar{\psi} = \bar{\psi} \left( \bar{I}_1 \, , \bar{I}_2 \, , 
\bar{I}_{\rm{4(\alpha \beta)}} \, , \bar{I}_{\rm{5(\alpha \beta)}} \right),
\end{aligned}
\label{eq:psibar}
\eeq
as functions of the isotropic and anisotropic deviatoric invariants from Eqs. \ref{eq:invisoF} and \ref{eq:invanisoF}, with $\alpha \, , \beta \in \{ 1,2,3 \}$ and $\beta \geq \alpha$.
\subsection{Constitutive neural network}
With the aim to universally model a hyperelastic history-independent soft matter material behavior,
we design a modular constitutive neural network architecture depicted in Figure \ref{figcnnarchitecture}.
Leveraging our prior work on automated constitutive model discovery for
isotropic \cite{Linka2023,Linka2023b,Peirlinck2024}, 
transversely isotropic \cite{Linka2023a,Peirlinck2023}, 
and orthotropic \cite{Martonova2024} soft materials,
we create a universal function approximator, which maps the 15 invariants $\bar{I}_1$, $\bar{I}_2$, $I_3$, $\bar{I}_{\rm{4(\alpha\beta)}}$, $\bar{I}_{\rm{5(\alpha\beta)}}$ of the deformation gradient $\ten{F}$ onto the free energy function $\psi \left( \ten{F} \right)$.
The constitutive relation between the Cauchy stress $\ten{\sigma}$ and the deformation gradient $\ten{F}$ follows naturally from the second law of thermodynamics as the derivative of the free energy function $\psi$ with respect to the deformation gradient $F$ according to Eq. \ref{eq:cauchyderivF}.
We ensure a vanishing free energy $\psi \left( \ten{F} \right) \doteq 0$ in the reference configuration, 
i.e., when $\ten{F}=\ten{I}$, 
by using the invariants' deviation from the energy-free reference state, 
$[\bar{I}_1-3]$,
$[\bar{I}_2-3]$,
$[I_3-1]$,
$[\bar{I}_{\rm{4(\alpha\beta)}}-\zeta_{\alpha\beta}]$,
$[\bar{I}_{\rm{5(\alpha\beta)}}-\zeta_{\alpha\beta}]$,
as constitutive neural network input.
Here, 
$\zeta_{\alpha\beta} = \vec{n}_{\alpha}^0 \cdot \vec{n}_{\beta}^0$ 
corrects invariants $\bar{I}_{\rm{4(\alpha\beta)}}$ and $\bar{I}_{\rm{5(\alpha\beta)}}$ for their values in the undeformed configuration.
This correction a priori ensures a \textit{stress-free reference configuration}.
To ensure polyconvexity, we design the constitutive neural network architecture as a locally connected, rather than a fully connected, feed forward neural network.
Specifically, we design the free energy function as a sum of individual polyconvex subfunctions with respect to each of the individual contributing invariants.
As a result, our free energy function from Eqs. \ref{eq:psiinvgeneral}-\ref{eq:psibar} can be  additively decomposed into
\beq
\begin{aligned}
\psi = 
&\bar{\psi}_1 (\bar{I}_1)
+ \bar{\psi}_2 (\bar{I}_2)
+ \psi_3 ({I}_3) + \\
& \sum_{\alpha=1}^N \sum_{\beta=\alpha}^{N} \bar{\psi}_{\rm{4 \left(\alpha\beta\right)}} \left(\bar{I}_{\rm{4\left(\alpha\beta\right)}}\right)+ \sum_{\alpha=1}^N \sum_{\beta=\alpha}^{N} \bar{\psi}_{\rm{5 \left(\alpha\beta\right)}} \left(\bar{I}_{\rm{5\left(\alpha\beta\right)}}\right),
\end{aligned}
\label{eq:psisummation}
\eeq
with $\alpha \, , \beta \in \{ 1,2,3 \}$ and $\beta \geq \alpha$. Following Eq. \ref{eq:cauchyderivF}, we derive the Cauchy stress
\beq
\begin{aligned}
 J \, \ten{\sigma}
&= 2 \frac{\partial \bar{\psi}_1}{\partial \bar{I}_1} \, \bar{\ten{b}}
  + 2 \frac{\partial \bar{\psi}_2}{\partial \bar{I}_2} \, [ \bar{I}_1 \bar{\ten{b}} - \bar{\ten{b}}^2 ] 
  + 2 \frac{\partial \psi_3}{\partial {I}_3} \, I_3 \ten{I} \\ 
  &+ \sum_{\alpha=1}^N \sum_{\beta=\alpha}^{N} \frac{\partial \bar{\psi}_{\rm{4(\alpha\beta)}}}{\partial \bar{I}_{\rm{4(\alpha\beta)}}} \, \left[ \bar{\vec{n}}_{\alpha} \otimes \bar{\vec{n}}_{\beta} + \bar{\vec{n}}_{\beta} \otimes \bar{\vec{n}}_{\alpha} \right] \\
  &+ \sum_{\alpha=1}^N \sum_{\beta=\alpha}^{N} \frac{\partial \bar{\psi}_{\rm{5(\alpha\beta)}}}{\partial \bar{I}_{\rm{5(\alpha\beta)}}} \, 
  \left[ 
  \bar{\vec{n}}_{\alpha} \otimes \bar{\ten{b}}\bar{\vec{n}}_{\beta} 
  + \bar{\ten{b}}\bar{\vec{n}}_{\alpha} \otimes \bar{\vec{n}}_{\beta}\right.\\ 
  &\hspace{2.8cm}+\left.\bar{\vec{n}}_{\beta} \otimes \bar{\ten{b}}\bar{\vec{n}}_{\alpha} 
  + \bar{\ten{b}}\bar{\vec{n}}_{\beta} \otimes \bar{\vec{n}}_{\beta} \right],
\end{aligned}
\label{eq:cauchystress}
\eeq
where $\bar{\vec{n}}_{\alpha} = \bar{\ten{F}} \cdot \vec{n}_{\alpha}^0$ and $\bar{\vec{n}}_{\beta} =  \bar{\ten{F}} \cdot \vec{n}_{\beta}^0$ represent the deviatoric fiber vectors in the current configuration.\\[6.pt] 
\noindent
Our constitutive network consists of three hidden layers with activation functions that are custom-designed to satisfy physically reasonable constitutive restrictions \cite{Antman2005,Linka2023}.
Specifically, we select from 
the identity $(\circ)$, 
the rectified linear unit function $\langle \circ \rangle$,
and the modulus function $| \circ |$ for the zeroth layer of the network,
from linear $(\circ)$,
quadratic $(\circ)^2$,
cubic $(\circ)^3$,
and higher order powers for the first layer,
and from linear $(\circ)$,
exponential $\exp(\circ)$,
and logarithmic $\ln(\circ)$ for the second layer.
\section{A universal material model}
\label{methods}
We incorporate our universal material model within a finite element analysis environment. Specifically, we develop a user-defined material model subroutine which functionally maps the local deformation gradient $\ten{F}$ onto the free energy function $\psi$ and computes its derivative with respect to the deformation gradient $\ten{F}$ and the Cauchy stress tensor $\ten{\sigma}$ using Eq. \ref{eq:cauchyderivF}. Additionally, we compute the stiffness tensor $\mathbb{C}$ to improve the accuracy, stability, and efficiency of the iterative solution technique required for an accurate prediction of the non-linear material behavior under various loading conditions.
\subsection{Subroutine}
To predict the quasi-static response of a system undergoing mechanical loading a non-linear finite element analysis solver iteratively evaluates whether a proposed update to the nodal displacement field satisfies the equilibrium equations that describe the force and momentum balance within the computational domain.
This evaluation requires the computation of the stress tensor and the tangent stiffness tensor as  functions of the proposed update to the body's total deformation.
At each time step, at each Newton-Raphson iteration, within each element, and for each integration point, the solver 
evaluates
the constitutive response that characterizes the functional mapping between the deformation gradient $\ten{F}$ and the stress tensor $\ten{\sigma}$ and tangent stiffness tensor $\mathbb{C}$.\\[6.pt]
%
\begin{figure}[t]
\centering
\includegraphics[width=1.0\linewidth]{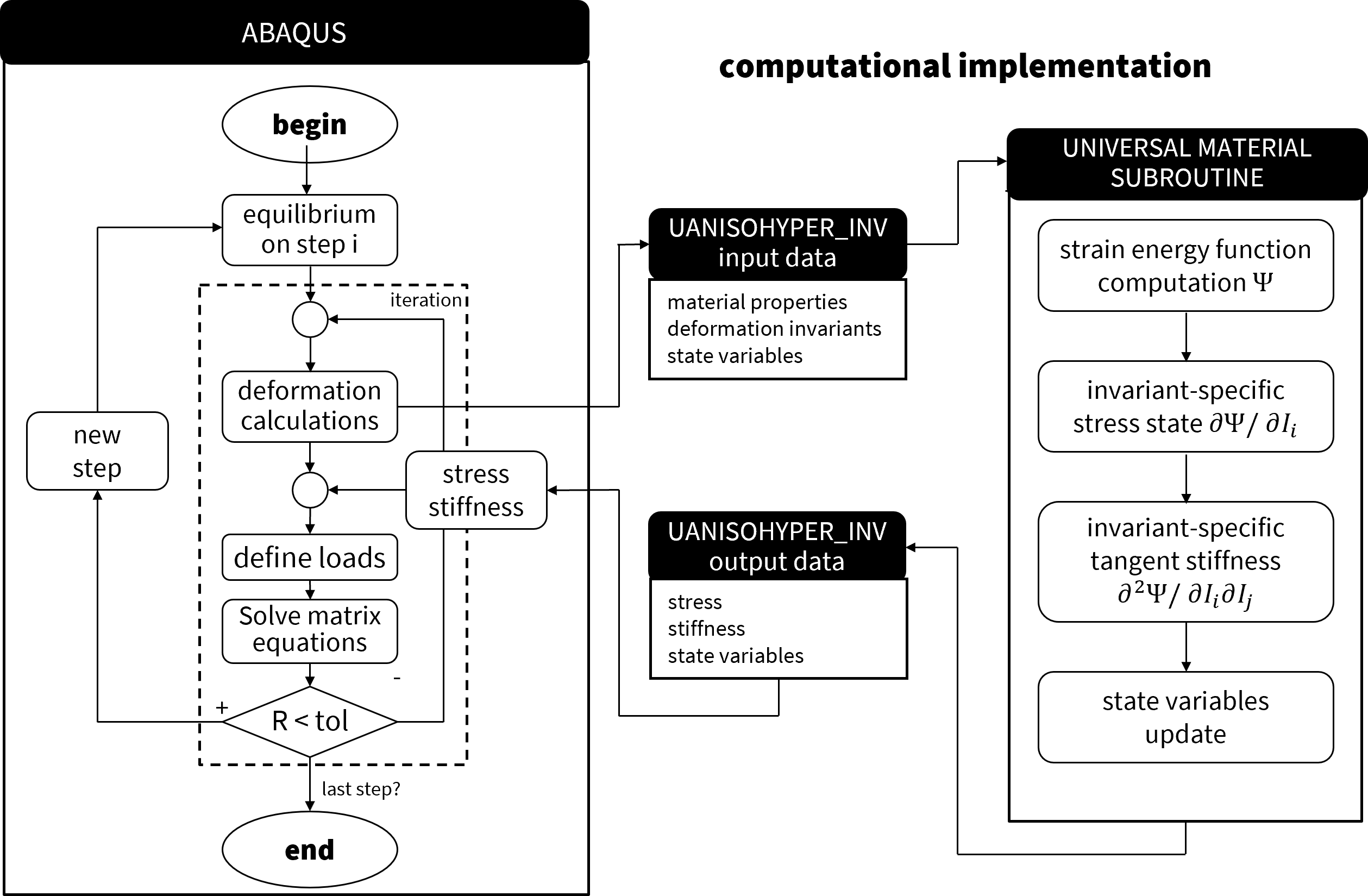}
\caption{{\textbf{\sffamily{Interaction between the finite element analysis solver and the universal material subroutine.}} Flowchart of the interaction between Abaqus and the \mbox{\texttt{UANISOHYPER\_INV}} subroutine architecture which embeds our universal constitutive material model. During each Newton-Raphson iteration and at each Gauss integration point, the \mbox{\texttt{UANISOHYPER\_INV}} subroutine computes the strain energy function $\psi$, its first derivatives with respect to the deformation invariants ${\partial \psi}/{\partial \bar{I}_i}$, and its second derivatives with respect to the deformation invariants ${\partial^2 \psi}/{\partial \bar{I}_i \partial \bar{I}_j}$. These quantities are used by Abaqus to compute the components of the Cauchy stress tensor and the material tangent stiffness tensor, to construct the element force vector and stiffness matrix, and to assemble the global righthand side vector and stiffness matrix. 
Abaqus then performs a Newton-Raphson iteration based on the residual between the internal and external forces, until it achieves convergence.}}
\label{figfeamatcallschematic}
\end{figure}
\begin{figure*}[t]
\centering
\includegraphics[width=0.9\linewidth]{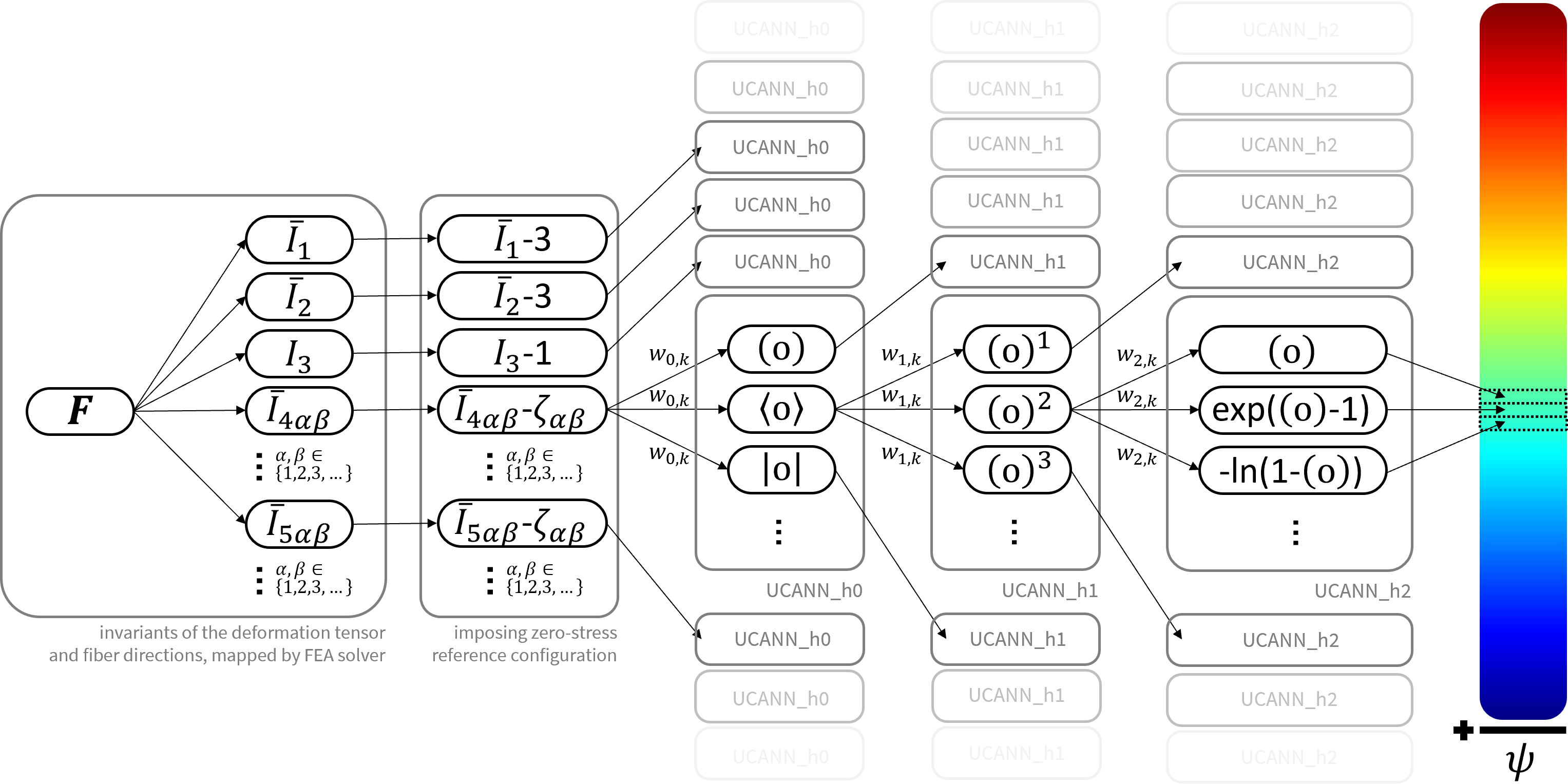}
\caption{{\textbf{\sffamily{Universal material model subroutine schematic.}} Our universal material model user subroutine computes the strain energy function $\psi$ (= \texttt{UA(1)}), 
its first derivatives ${\partial \psi}/{\partial I_i}$ (= \texttt{UI1(NINV)}),
and its second derivatives ${\partial^2 \psi}/{\partial \bar{I}_i \partial \bar{I}_j}$ (=\texttt{UI2(NINV$^{*}$(NINV+1)/2)})
with respect to the scalar invariants $\bar{I}_i$, derived from the deformation gradient $\ten{F}$.
These functions and derivatives are computed based on a triple set of nested activation functions
$f_{0}$ (= \texttt{UCANN\_h0}),
$f_{1}$ (= \texttt{UCANN\_h1}), and
$f_{2}$ (= \texttt{UCANN\_h2}),
where each unique constitutive path forms an additive constitutive \textit{neuron} towards the total free energy and its derivatives.
}}\label{figuanisoschematic}
\end{figure*}\\
We leverage the \mbox{\texttt{UANISOHYPER\_INV}} user-defined subroutine architecture in Figure \ref{figfeamatcallschematic} to seamlessly integrate our universal constitutive neural network architecture 
within the Abaqus finite element analysis software suite \cite{Abaqus2024}.
This subroutine provides three input arrays: the material properties we provide in the finite element analysis input file; the deformation gradient invariants as defined in Eqs. \ref{eq:inv3F},\ref{eq:invisoF},\ref{eq:invanisoF}, and \ref{eq:invanisocoupleF}; and an array of state-dependent field variables.
Upon each evaluation, our user-defined subroutine \textit{updates}
the free energy function \texttt{UA(1)} $= \psi$,
the array of first derivatives of the free energy with respect to the scalar invariants
\texttt{UI1(NINV)} $= {\partial \psi}/{\partial \bar{I}_i}$, 
and the array of second derivatives of the free energy with respect to the scalar invariants
\texttt{UI2(NINV$^{*}$(NINV+1)/2)}$= {\partial^2 \psi}/{\partial \bar{I}_i \partial \bar{I}_j}$.\\[6.pt]
Figure \ref{figuanisoschematic} showcases the internal code structure of our universal material model subroutine.
We construct a triple set of nested activation functions 
\mbox{\texttt{UCANN\_h0}},  \mbox{\texttt{UCANN\_h1}}, and \mbox{\texttt{UCANN\_h2}} 
to compute $\psi$, ${\partial \psi}/{\partial \bar{I}_i}$, ${\partial^2 \psi}/{\partial \bar{I}_i \partial \bar{I}_j}$ from $\bar{I}_i$. 
We adopt the invariant numbering,
\beq
\begin{array}{lcl}
 \bar{I}_{1} &\rightarrow& \bar{I}_{\texttt{NINV}}; \; \texttt{NINV} = 1 \\
 \bar{I}_{2} &\rightarrow& \bar{I}_{\texttt{NINV}}; \; \texttt{NINV} = 2 \\
 I_{3} &\rightarrow& I_{\texttt{NINV}}; \; \texttt{NINV} = 3 \\
 \bar{I}_{4(\alpha\beta)} 
       &\rightarrow& \bar{I}_{\texttt{NINV}}; \; \texttt{NINV} = 4 + 2 \, (\alpha-1)+\beta \, (\beta-1))  \\
 \bar{I}_{5(\alpha\beta)} 
       &\rightarrow& \bar{I}_{\texttt{NINV}}; \; \texttt{NINV} = 5 + 2 \, (\alpha-1)+\beta \, (\beta-1))
\end{array}
\label{eq:ninv}
\eeq
Dependent on the number of fiber families, this scheme automatically adapts itself to account for multiple fiber orientations. For example, when our material displays an anisotropic behavior with three families of fibers (${\tt{NDIR}=3}$), there are a total of 15 invariants: $\bar{I}_{1}$, $\bar{I}_{2}$, $I_{3} = J$, six invariants of type $\bar{I}_{4(\alpha\beta)}$, and six invariants of type $\bar{I}_{5(\alpha\beta)}$, 
with $\alpha \, , \beta \in \{ 1,2,3 \}$ and $\beta \geq \alpha$.
\paragraph{Free energy function update.}
Without loss of generality, we reformulate the free energy $\psi$ from Eq. \ref{eq:psisummation} in the following form,
\beq
\begin{aligned}
\psi    &= f_{2} \circ  f_{1} \circ f_{0} \, (\bar{I}_{i}-\bar{I}_{i0}) \\
        &= \sum_{k=1}^n w_{2,k} \, f_{2,k} \, (f_{1,k} \, (f_{0,k} \,(\bar{I}_{i,k}-\bar{I}_{i0,k}); w_{1,k})) \,,
\end{aligned}
\label{eq:UNAIenergynested}
\eeq
where $f_{0}$, $f_{1}$, $f_{2}$ are the nested activation functions associated with the zeroth, first, and second layers of our constitutive neural network; 
$k=1,...,n$ defines each unique additive constitutive \textit{neuron} that stems from the expanding nested constitutive neural network in Figure \ref{figuanisoschematic};
and $\bar{I}_{i0}$ imposes the free energy $\psi$ and Cauchy stress $\ten{\sigma}$ to be zero in the reference configuration. As discussed above and shown in Figure \ref{figcnnarchitecture}, these corrections amount to 
$\bar{I}_{i0}=3$ for $i=1,2$, 
to $I_{i0}=1$ for $i=3$, 
and $\bar{I}_{i0}=\zeta_{\alpha\beta} = \vec{n}_{\alpha}^0 \cdot \vec{n}_{\beta}^0$ for $i \geq 4$ 
with respect to the invariant numbering scheme in Eq. \ref{eq:ninv}.
Our nested activation functions in Eq. \ref{eq:UNAIenergynested} read
\beq  
  f_0 = \left\{
  \begin{array}{c}
  {\left(\circ\right)}\\
  {\langle\circ\rangle}\\
  {\lvert\circ\rvert}\\
  \vdots 
  \end{array} \right.
  \;
  f_1 = \left\{
  \begin{array}{c}
  {(\circ)^1}\\
  {(\circ)^2}\\
  {(\circ)^3}\\
  \vdots \\
  {(\circ)^m}
  \end{array} \right.
  \;
  f_2 = \left\{
  \begin{array}{c}
  {w_1 \, (\circ)}\\
  {{\rm{exp}}(w_1 \, (\circ))-1}\\
 -{{\rm{ln}}(1-w_1 \, (\circ))} \\
  \vdots
  \end{array} \right. \!\!.
  \label{eq:UNAIenergynested}
\eeq
The activation function
$f_0$ returns the identity, Macauley bracketed, or absolute values, $(\circ)$, $\langle \circ \rangle$, $\lvert \circ \rvert$ of the zero-stress reference configuration corrected invariants;
$f_1$ raises these invariants to the first, second, third, or any higher order powers, 
$(\circ)^1$, $(\circ)^2$, $(\circ)^3$, $(\circ)^m$; 
and $f_2$ applies the identity, exponential, or natural logarithm, 
$(\circ)$, $(\rm{exp}(\circ)-1)$, $(-\rm{ln}(1-(\circ)))$, 
or any other thermodynamically admissible function to these powers.\\[6.pt]
\paragraph{Cauchy stress tensor update.}
To update the Cauchy stress tensor $\ten{\sigma}$, we reformulate Eq. \ref{eq:cauchyderivF} in the following form,
\beq
\begin{aligned}
\ten{\sigma}&=\frac{1}{J}\frac{\partial \psi\left( \ten{F} \right)}{\partial \ten{F}} \cdot \ten{F}^{\scas{t}}
=\sum_{k=1}^n \frac{1}{J}\frac{\partial \psi}{\partial \bar{I}_{i,k}}\frac{\partial \bar{I}_{i,k}}{\partial \ten{F}} \cdot \ten{F}^{\scas{t}}\\
&=\sum_{i} \frac{1}{J} \left(\sum_{k}^{n_i}\frac{\partial \psi}{\partial \bar{I}_{i,k}}\right)\frac{\partial \bar{I}_{i}}{\partial \ten{F}} \cdot \ten{F}^{\scas{t}}
=\sum_{i} \frac{1}{J} \frac{\partial \psi}{\partial \bar{I}_{i}}\frac{\partial \bar{I}_{i}}{\partial \ten{F}} \cdot \ten{F}^{\scas{t}}\\
\end{aligned}
\label{eq:cauchyderivFsplitted}
\eeq
where Abaqus internally computes of the ${\partial \bar{I}_{i}}/{\partial \ten{F}}$ terms and the summation of the individual \texttt{NINV} stress tensor contributions. 
We compute all the invariant-specific scalar \texttt{UI1(NINV)} $= {\partial \psi}/{\partial \bar{I}_i}$ contributions to the full \texttt{UI1} array in our user subroutine,
\beq
\frac{\partial \psi}{\partial \bar{I}_{i}} 
= \sum_{k}^{n_i}\frac{\partial \psi}{\partial \bar{I}_{i,k}}
= \sum_{k}^{n_i} w_{2, \mathrm{k}} \frac{\partial f_{2,k}}{\partial(\circ)} \frac{\partial f_{1,k}}{\partial(\circ)} \frac{\partial f_{0,k}}{\partial \bar{I}_{i,k}}
\eeq
in terms of the first derivatives of our activation functions
\beq
  \frac{\partial f_{0}}{\partial (\circ)} = \left\{
  \begin{array}{@{\hspace*{0.0cm}}c}
  1\\
  \frac{\frac{\lvert \circ \rvert}{\circ} + 1}{2}\\ 
  \frac{\lvert \circ \rvert}{\circ}\\ 
  \vdots
  \end{array} \right.
  \frac{\partial f_{1}}{\partial (\circ)} = \left\{
  \begin{array}{@{\hspace*{0.0cm}}c}
  {1 (\circ)^0}\\
  {2 (\circ)^1}\\
  {3 (\circ)^2}\\
  \vdots \\
  m (\circ)^{m-1}
  \end{array} \right. \!\!
  \frac{\partial f_{2}}{\partial (\circ)} = \left\{
  \begin{array}{@{\hspace*{0.0cm}}c}
  {w_1}\\
  {w_1  {\rm{exp}}(w_1 (\circ))}\\
  {w_1  /(1\mbox{-}w_1 (\circ))} \\
  \vdots
  \end{array} \right.
  \label{FEMstress}
\eeq
\paragraph{Tangent stiffness tensor update.}
To internally compute the tangent stiffness tensor, Abaqus needs the second derivatives of the strain energy function with respect to the invariants ${\partial^2 \Psi}/{\partial \bar{I}_{i,k} \partial \bar{I}_{j,k}}$. 
Here, given the nested structure of our universal material model subroutine, we have ${\partial^2 \Psi}/{\partial \bar{I}_{i,k} \partial \bar{I}_{j,k}} = 0$ , when $i \neq j$ 
As such, we only have non-zero values

\beq
\begin{aligned}
\frac{\partial^2 \Psi}{\partial \bar{I}_{i,k}^2}=\sum_{k=1}^{n_i} w_{2,k}
& \left[\left(\frac{\partial^2 f_{2,k}}{\partial(\circ)^2}\left[\frac{\partial f_{1,k}}{\partial(\circ)}\right]^2+\frac{\partial f_{2,k}}{\partial(\circ)} \frac{\partial^2 f_{1,k}}{\partial(\circ)^2}\right) \right.\\
& \left.\hspace{0.4cm}\left[\frac{\partial f_{0,k}}{\partial \bar{I}_{i,k}}\right]^2+\frac{\partial f_{2,k}}{\partial(\circ)} \frac{\partial f_{1,k}}{\partial(\circ)} \frac{\partial^2 f_{0,k}}{\partial \bar{I}_{i,k}^{2}} \right]
\end{aligned}
\eeq
in terms of the second derivatives of our activation functions,

\beq
\begin{array}{l@{\hspace{-0.2cm}}l}
\D{\frac{\partial^2 f_1}{\partial \left(\circ\right)^2}} = \left\{
  \begin{array}{c}
  0\\
  2\\
  6\\
  \vdots \\
  (m^2 -m) (\circ)^{m-2}
  \end{array} \right.
&
\D{\frac{\partial^2 f_2}{\partial \left(\circ\right)^2}} = \left\{
    \begin{array}{c}
    0\\
    {{w_1}^2{\rm{exp}}(w_1(\circ))}\\
    {{w_1}^2/{(1-w1(\circ))^2}} \\
    \vdots 
    \end{array} \right. 
\end{array}
\label{eq:UNAIenergynested}
\eeq
where the second derivative of the zeroth layer functions, $\partial^2 f_0/\partial(\circ)^2$, vanishes identically for all three terms. 
\subsection{Pseudocodes}
In the following five algorithmic boxes, we summarize our universal material subroutine as pseudocode.\\[6.pt] 
\begin{algorithm}
\SetKwFunction{uanisohyperinv}{\texttt{UANISOHYPER\_INV}}
\SetKwFunction{uCANN}{uCANN}
\caption{Pseudocode for universal material subroutine \texttt{UANISOHYPER\_INV}}
\SetKwProg{sub}{subroutine}{}{}
\sub{\uanisohyperinv{{\emph{\texttt{aInv,UA,UI1,UI2}}}}}{
    \BlankLine
    \tcp{initialize variables}
    {\texttt{set initial array values for}} \texttt{UA}, \texttt{UI1}, \texttt{UI2}\;
    {\texttt{set reference configuration}} \texttt{UANISOHYPER\_INV}\;
    {\texttt{set discovered parameters}} \texttt{UNIVERSAL\_TAB}\;
    \BlankLine
    \tcp{evaluate all n rows in parameter table}
    \For{{\emph{\texttt{k in n}}}}{
        \tcp{invariant,\,activation\,functions,\,weights}
        {\texttt{extract invariant}} {\texttt{kf0(k)\;}}
        {\texttt{extract\,activation\,functions}}\,{\texttt{kf1(k)}},\,{\texttt{kf2(k)}};
        {\texttt{extract weights}} {\texttt{w1(k)}}, {\texttt{w2(k)\;}}
        \BlankLine
        \tcp{invariant in reference configuration}
        {\texttt{xInv = aInv(kf0(k))-xInv0(k)\;}}
        \BlankLine
        \tcp{energy\,and\,derivatives \texttt{UA},\,\texttt{UI1},\,\texttt{UI2}}
        \BlankLine
        {\texttt{call}} \uCANN{{\emph{\texttt{xInv,kf1(k),kf2(k),w1(k),}}}
        \hspace*{1.66cm}{\emph{\texttt{w2(k),UA,UI1,UI2}}}};
    }
    \BlankLine
    \tcp{return updated arrays}
    \textbf{return} \texttt{UA}, \texttt{UI1}, \texttt{UI2}
}
\label{algr01}
\end{algorithm}

\begin{algorithm}
\SetKwFunction{uCANN}{uCANN}
\SetKwFunction{uCANNhzero}{uCANN\_h0}
\SetKwFunction{uCANNhone}{uCANN\_h1}
\SetKwFunction{uCANNhtwo}{uCANN\_h2}
\SetKwProg{sub}{subroutine}{}{}
\caption{Pseudocode to update energy and its derivatives \texttt{UA, UI1, UI2}}
\sub{\uCANN{\emph{\texttt{xInv,kf1,kf2,w1,w2,UA,UI1,UI2}}}}{ 
    \BlankLine
    \tcp{zeroth layer: calculate {\texttt{f0,df0,ddf0}}}
    {\texttt{call}}  \uCANNhzero{\emph{\texttt{xInv,kf0,f0,df0,ddf0}}}\;
    \BlankLine
    \tcp{first layer: calculate {\texttt{f1,df1,ddf1}}}
    {\texttt{w0 = 1\;}}
    {\texttt{call}}  \uCANNhone{\emph{\texttt{xInv,w0,kf1,f1,df1,ddf1}}}\;
    \BlankLine
    \tcp{second layer: calculate {\texttt{f2,df2,ddf2}}}
    {\texttt{call}} \uCANNhtwo{\emph{\texttt{f1,w1,kf2,f2,df2,ddf2}}}\;
    \BlankLine
    \tcp{update energy and derivatives \texttt{UA},\texttt{UI1},\texttt{UI2}}
    {\texttt{UA \hspace*{0.18cm}= UA  \hspace*{0.18cm}+ w2 *  f2\;}}
    {\texttt{UI1 = UI1 + w2 * df2*df1*df0\;}}
    {\texttt{UI2 = UI2 + w2 *((ddf2*df1**2+df2*ddf1)}}
    {\texttt{\hspace*{2.8cm}*df0**2+df2*df1*ddf0)\;}}
    {\textbf{return}} \texttt{UA}, \texttt{UI1}, \texttt{UI2}
}
\label{algr02}
\end{algorithm} 

\begin{algorithm}
\SetKwFunction{uCANN}{uCANN}
\SetKwFunction{uCANNhzero}{uCANN\_h0}
\SetKwFunction{uCANNhone}{uCANN\_h1}
\SetKwFunction{uCANNhtwo}{uCANN\_h2}
\SetKwProg{sub}{subroutine}{}{}
\caption{Pseudocode to evaluate output of zeroth network layer {\texttt{f,df,ddf}}}
\sub{\uCANNhzero{\emph{\texttt{x,kf,f,df,ddf}}}}{
    \tcp{calculate zero layer output {\texttt{f,df,ddf}} for activation function {\texttt{kf}}}
    \BlankLine
    \uIf{\emph{\texttt{ kf = 1 }}}{
    {\texttt{f = x\;}}
    {\texttt{df = 1\;}}
    {\texttt{ddf = 0\;}}
    }
    \BlankLine
    \uElseIf{\emph{\texttt{kf = 2}}}{
    {\texttt{f = $(\lvert x \rvert+x)/2$\;}}
    {\texttt{df = $(\lvert x \rvert / x + 1)/2$\;}}
    {\texttt{ddf = 0\;}}
    }
    \BlankLine
    \uElseIf{\emph{\texttt{kf = 3}}}{
    {\texttt{f = $\lvert x \rvert$\;}}
    {\texttt{df = $\lvert x \rvert/x$\;}}
    {\texttt{ddf = 0\;}}
    }
    \BlankLine
    {\bf{return}} {\texttt{f,df,ddf}}
}
\label{algr03}
\end{algorithm}

\begin{algorithm}
\SetKwFunction{uCANN}{uCANN}
\SetKwFunction{uCANNhone}{uCANN\_h1}
\SetKwFunction{uCANNhtwo}{uCANN\_h2}
\SetKwProg{sub}{subroutine}{}{}
\caption{Pseudocode to evaluate output of first network layer {\texttt{f,df,ddf}}}
\sub{\uCANNhone{\emph{\texttt{x,w,kf,f,df,ddf}}}}{
    \tcp{calculate first layer output {\texttt{f,df,ddf}} for activation function {\texttt{kf}}}
    \BlankLine
    \uIf{\emph{\texttt{ kf = 1 }}}{
    {\texttt{f = w * x\;}}
    {\texttt{df = w * 1\;}}
    {\texttt{ddf = w * 0\;}}
    }
    \BlankLine
    \uElseIf{\emph{\texttt{kf = 2}}}{
    {\texttt{f = w**2 * x**2\;}}
    {\texttt{df = w**2 * 2*x\;}}
    {\texttt{ddf = w**2 * 2\;}}
    }
    \BlankLine
    {\bf{return}} {\texttt{f,df,ddf}}
}
\label{algr04}
\end{algorithm}

\begin{algorithm}
\SetKwFunction{uCANN}{uCANN}
\SetKwFunction{uCANNhone}{uCANN\_h1}
\SetKwFunction{uCANNhtwo}{uCANN\_h2}
\SetKwProg{sub}{subroutine}{}{}
\caption{Pseudocode to evaluate output of second network layer {\texttt{f,df,ddf}}}
\sub{\uCANNhtwo{\emph{\texttt{x,kf,w,f,df,ddf}}}}{
    \tcp{calculate second layer output {\texttt{f,df,ddf}} for activation function {\texttt{kf}}}
    \BlankLine
    \uIf{\emph{\texttt{ kf = 1 }}}{
    {\texttt{f = w * x\;}}
    {\texttt{df = w * 1\;}}
    {\texttt{ddf = w * 0\;}}
    }
    \BlankLine
    \uElseIf{\emph{\texttt{ kf = 2 }}}{
    {\texttt{f = exp(w*x)-1\;}}
    {\texttt{df = w * exp(w*x)\;}}
    {\texttt{ddf = w**2 * exp(w*x)\;}}
    }
    \BlankLine
    \uElseIf{\emph{\texttt{ kf = 3 }}}{
    {\texttt{f = -ln(1-w*x)\;}}
    {\texttt{df = w / (1-w*x)\;}}
    {\texttt{ddf = w**2 / (1-w*x)**2\;}}
    }
    \BlankLine
    {\bf{return}} {\texttt{f,df,ddf}}
}
\label{algr05}
\end{algorithm}
\noindent Algorithm \ref{algr01} illustrates the \texttt{UANISOHYPER\_INV} pseudocode to compute the arrays \texttt{UA(1)}, \texttt{UI1(NINV)}, and \texttt{UI2}\texttt{(NINV} \texttt{$^*$(NINV+1)/2)} at the integration point level. 
First, we initialize all relevant arrays and read the activation functions $kf_{0,k}$, $kf_{1,k}$ and $kf_{2,k}$ and weights $w_{0,k}$, $w_{1,k}$ and $w_{2,k}$ of the $n$ constitutive neurons of our constitutive neural network from our user-defined parameter table \texttt{UNIVERSAL\_TAB}, where $w_{0,k}=1.0$ by default.
Then, for each node, we evaluate its row in the parameter table \texttt{UNIVERSAL\_TAB} and additively update the strain energy density function and its first and second derivatives,
\texttt{UA}, \texttt{UI1}, \texttt{UI2}. \\[6.pt]
\noindent Algorithm \ref{algr02} summarizes the additive update of the free energy and its first and second derivatives, \texttt{UA}, \texttt{UI1}, \texttt{UI2}, within the universal material subroutine \texttt{uCANN}. \\[6.pt] 
\noindent Algorithms \ref{algr03},\ref{algr04} and \ref{algr05} provide the pseudocode for the three 
subroutines \mbox{\texttt{uCANN\_h0}}, \mbox{\texttt{uCANN\_h1}} and \mbox{\texttt{uCANN\_h2}} 
that evaluate the zeroth, first and second network layers for each network node with its discovered activation functions and weights. \\[6.pt]
\subsection{FEA integration}
The concept of our universal material subroutine is inherently modular and generally compatible with any finite element analysis package \cite{Abaqus2024,Logg2012,Maas2012,Updegrove2016,Africa2022}. 
Here, for illustrative purposes, we implement the universal material subroutine in Abaqus FEA \cite{Abaqus2024}, and make all our code and simulation files publicly available on Zenodo. 
For the input file to our finite element simulation, 
we define our discovered model and parameters in a parameter table. 
Each row of this table represents a 
neuron of the final layer in our constitutive neural network
and consists of seven terms:
an integer {\tt{kfinv}} that defines the index of the invariant $\bar{I}_i$ according to the invariant numbering scheme in equation \ref{eq:ninv};
three integers {\tt{kf0}},{\tt{kf1}} and {\tt{kf2}} that define the indices of the zeroth-, first-, and second-layer activation functions; and
three floats {\tt{w0}}, {\tt{w1}} and {\tt{w2}} that define the weights of the zeroth, first, and second layers.
In Abaqus,
we declare the format of this input parameter table using the parameter table type definition in the \mbox{\tt{UNIVERSAL\_PARAM\_TYPES.INC}} file.
\\[6.pt]
\begin{small}
$\begin{array}{@{}l}
\mbox{{\bf{\tt{*PARAMETER TABLE TYPE, name="UNIVERSAL\_TAB",}}}} \\ 
\hspace*{0.2cm} \mbox{{\bf{\tt{parameters = 7}}}} \\
\begin{array}{lll}
\mbox{\bf{\tt{INTEGER}}}\,, &\,\bf{\tt{,}}&\mbox{\bf{\tt{"index  invariant, kfinv,o"}}} \\
\mbox{\bf{\tt{INTEGER\,,}}} &\,\bf{\tt{,}}&\mbox{\bf{\tt{"index  0th activ function, kf0,o"}}} \\
\mbox{\bf{\tt{INTEGER\,,}}} &\,\bf{\tt{,}}&\mbox{\bf{\tt{"index  1st activ function, kf1,o"}}} \\
\mbox{\bf{\tt{INTEGER\,,}}} &\,\bf{\tt{,}}&\mbox{\bf{\tt{"index  2nd activ function, kf2,o"}}} \\
\mbox{\bf{\tt{FLOAT\,,}}} &\,\bf{\tt{,}}&\mbox{\bf{\tt{"weight 0th hidden layer, w0,o"}}} \\
\mbox{\bf{\tt{FLOAT\,,}}} &\,\bf{\tt{,}}&\mbox{\bf{\tt{"weight 1st hidden layer, w1,o"}}} \\
\mbox{\bf{\tt{FLOAT\,,}}} &\,\bf{\tt{,}}&\mbox{\bf{\tt{"weight 2nd hidden layer, w2,o"}}} \\
\end{array} 
\end{array}$
\end{small}
\\[6.pt]
Within our Abaqus FEA simulation input file, we include the parameter table type definition using 
\\[6.pt]
\begin{small}
$\begin{array}{@{}l}
\mbox{{\bf{\tt{*INCLUDE, INPUT=UNIVERSAL\_PARAM\_TYPES.INC}}}}
\end{array} $
\end{small}
\\[6.pt]
and call our user-defined material model through the command 
\\[6.pt]
\begin{small}
$\begin{array}{@{}l}
\mbox{{\bf{\tt{*ANISOTROPIC\;HYPERELASTIC,\,USER,\,FORMULATION=INVARIANT,}}}} \\
\hspace*{0.15cm}\mbox{{\bf{\tt{TYPE=INCOMPRESSIBLE, LOCAL DIRECTIONS=NDIR}}}}
\end{array}$ 
\end{small}
\\[6.pt]
where integer {\tt{NDIR}} defines the number of local fiber directions in our material,
followed by the discovered list of parameters:
\\[6.pt]
\begin{small}
$\begin{array}{@{}l}
\mbox{{\bf{\tt{*PARAMETER TABLE, TYPE="UNIVERSAL\_TAB"}}}} \\
\mbox{{\bf{\tt{1,1,1,1,1.0,w$_{\tt{1,1}}$,w$_{\tt{2,1}}$}}}} \\ 
\mbox{{\bf{\tt{1,1,1,2,1.0,w$_{\tt{1,2}}$,w$_{\tt{2,2}}$}}}} \\ 
\mbox{{\bf{\tt{1,1,2,1,1.0,w$_{\tt{1,3}}$,w$_{\tt{2,3}}$}}}} \\ 
\mbox{{\bf{$\vdots$}}}
\end{array}$ 
\end{small}
\\[6.pt]
The first index of each row selects between the invariants, 
the second index applies the identity, Macauley brackets, or absolute values to the invariants, 
$(\circ)$, $\langle \circ \rangle$, $\lvert \circ \rvert$,
the third index raises them the first, second, third, or any higher order powers, 
$(\circ)^1$, $(\circ)^2$, and
the fourth index applies the identity, exponential, or natural logarithm, 
$(\circ)$, $(\rm{exp}(\circ)-1)$, $(-\rm{ln}(1-(\circ)))$, 
or any other thermodynamically admissible function to these powers.
For brevity, we can simply exclude terms with zero weights from the list. 
\subsection{Compressibility}
To extend our universal material model subroutine towards compressible materials, we add a volumetric strain energy function $\psi_{\rm{vol}}$ in terms of the third invariant $I_3$. We follow the same nested activation function approach as in Figure \ref{figuanisoschematic}, and  add the volumetric strain energy density $\psi^{\rm{vol}}$, 
its first derivative ${\partial \psi^{\rm{vol}}}/{\partial I_3}$ 
and its second derivatives ${\partial^2 \psi_{\rm{vol}}}/{\partial I_3^2}$ 
to the \texttt{UA(1)}, \texttt{UI1(NINV)}, \texttt{UI2(NINV$^{*}$(NINV+1)/2)} arrays.
Using the invariant numbering scheme from equation \ref{eq:ninv}, we have NINV$=3$, and  introduce \texttt{UA(1)}, \texttt{UI1(3)} and \texttt{UI2(6)}.
%
%
To incorporate compressible material behavior in our FEA simulation,
we change the \texttt{TYPE} keyword argument line in our Abaqus input file to \texttt{TYPE = COMPRESSIBLE}, 
\\[6.pt]
\begin{small}
$\begin{array}{@{}l}
\mbox{{\bf{\tt{*ANISOTROPIC\;HYPERELASTIC,\,USER,\,FORMULATION=INVARIANT,}}}} \\
\hspace*{0.15cm}\mbox{{\bf{\tt{TYPE=COMPRESSIBLE, LOCAL DIRECTIONS=NDIR}}}}
\end{array}$ 
\end{small}
\\[6.pt]
where integer {\tt{NDIR}} defines the number of local fiber directions of our material.
We add the volumetric contributions to our constitutive parameter table, along with all the other deviatoric free energy contributions, 
\\[6.pt]
\begin{small}
$\begin{array}{@{}l}
\mbox{{\bf{\tt{*PARAMETER TABLE, TYPE="UNIVERSAL\_TAB"}}}} \\
\begin{array}{ttttttt}
    \vdots &\vdots &\vdots &\vdots &\vdots &\vdots & \vdots \\
     3,&1,&1,&1,&1.0,&w_{1,\circ},& w_{2,\circ} \\
     3,&1,&2,&1,&1.0,&w_{1,\circ},& w_{2,\circ}
\end{array}
\end{array}$ 
\end{small}
\\[12.pt]
\noindent
For example, the volumetric strain energy function \cite{Simo1988}
\beq
\psi_{\mathrm{vol}}=\frac{K}{2}(I_3 - 1)^2
\label{volSEFsimo1988}
\eeq
translates into the following contribution to the input file
\\[6.pt]
\begin{small}
$\begin{array}{@{}l}
\mbox{{\bf{\tt{*PARAMETER TABLE, TYPE="UNIVERSAL\_TAB"}}}} \\
\mbox{{\bf{\tt{3,1,2,1,1.0,1.0,K/2}}}} \\ 
\end{array}$ 
\end{small}
\\[12.pt]
\noindent
Alternatively, the volumetric strain energy function
\beq
\psi_{\mathrm{vol}}=\frac{K}{2} \left( \frac{I_3^2 - 1}{2} - \ln(I_3) \right)
\label{volSEFogden1972}
\eeq
which is a special case of the modified Ogden formulation \cite{Ogden1972},
can be reformulated to
\begin{equation}
\psi_{\mathrm{vol}} 
= \frac{K}{2}\left( \!(I_3-1) + \frac{1}{2}(I_3-1)^2 - \ln (1-(-1)(I_3-1)) \!\right)
\label{volSEFogden1972alt}
\end{equation}
which translates into the following lines in the input file
\\[6.pt]
\begin{small}
$\begin{array}{@{}l}
\mbox{{\bf{\tt{*PARAMETER TABLE, TYPE="UNIVERSAL\_TAB"}}}} \\
\begin{array}{ttttttt}
     3,&1,&1,&1,&1.0,&1.0,&K/2 \\
     3,&1,&2,&1,&1.0,&0.5,&K/2 \\
     3,&1,&1,&3,&1.0,&\text{-}1.0,&K/2
\end{array}
\end{array}$ 
\end{small}
\\[6.pt]
\subsection{The subroutine applied}
To showcase the flexibility and modularity of our universal material model subroutine, we demonstrate how our approach naturally integrates the popular
neo Hooke \cite{Treloar1948}, 
Mooney Rivlin \cite{Mooney1940,Rivlin1948},
Yeoh \cite{Yeoh1993}, 
polynomial \cite{Rivlin1951}, 
Holzapfel \cite{Holzapfel2000}, and 
Kaliske \cite{Kaliske2004}
models into Abaqus FEA. For each model, we provide the strain energy function and its translation into the \texttt{UNIVERSAL\_TAB} parameter table for the FEA input file.
\paragraph{Neo Hooke model}
The strain energy function of the compressible linear first invariant neo Hooke model \cite{Treloar1948}
\begin{equation}
\psi=C_{10}\left(\bar{I}_1-3\right)+\frac{1}{D_1}\left(I_3-1\right)^2
\end{equation}
translates into the following two-line parameter table
\\[6.pt]
\begin{small}
$\begin{array}{@{}l}
\mbox{{\bf{\tt{*PARAMETER TABLE, TYPE="UNIVERSAL\_TAB"}}}} \\
\begin{array}{ttttttt}
     1,&1,&1,&1,&1.0,&1.0,& C_{10} \\
     3,&1,&2,&1,&1.0,&1.0,& 1/D_{1}
\end{array}
\end{array}$ 
\end{small}
\\[6.pt]
\paragraph{Mooney Rivlin model}
The strain energy function of the compressible linear first and second invariant Mooney Rivlin model \cite{Mooney1940,Rivlin1948}
\begin{equation}
\psi = C_{10}\left(\bar{I}_1-3\right)+C_{01}\left(\bar{I}_2-3\right)+\frac{1}{D_1}\left(I_3-1\right)^2
\end{equation}
translates into the following three-line parameter table
\\[6.pt]
\begin{small}
$\begin{array}{@{}l}
\mbox{{\bf{\tt{*PARAMETER TABLE, TYPE="UNIVERSAL\_TAB"}}}} \\
\begin{array}{ttttttt}
     1,&1,&1,&1,&1.0,&1.0,& C_{10} \\
     2,&1,&1,&1,&1.0,&1.0,& C_{01} \\
     3,&1,&2,&1,&1.0,&1.0,& 1/D_{1}
\end{array}
\end{array}$ 
\end{small}
\\[6.pt]
\paragraph{Yeoh model}
The strain energy function of the compressible first invariant Yeoh model \cite{Yeoh1993}
\begin{equation}
\begin{aligned}
\psi & =  C_{10}\left(\bar{I}_1-3\right)+C_{20}\left(\bar{I}_1-3\right)^2+C_{30}\left(\bar{I}_1-3\right)^3 \\
& +\frac{1}{D_1}\left(I_3-1\right)^2+\frac{1}{D_2}\left(I_3-1\right)^4+\frac{1}{D_3}\left(I_3-1\right)^6
\end{aligned}
\end{equation}
translates into the following six-line parameter table
\\[6.pt]
\begin{small}
$\begin{array}{@{}l}
\mbox{{\bf{\tt{*PARAMETER TABLE, TYPE="UNIVERSAL\_TAB"}}}} \\
\begin{array}{ttttttt}
     1,&1,&1,&1,&1.0,&1.0,& C_{10} \\
     1,&1,&2,&1,&1.0,&1.0,& C_{20} \\
     1,&1,&3,&1,&1.0,&1.0,& C_{30} \\
     3,&1,&2,&1,&1.0,&1.0,& 1/D_{1} \\
     3,&1,&4,&1,&1.0,&1.0,& 1/D_{2} \\
     3,&1,&6,&1,&1.0,&1.0,& 1/D_{3} \\
\end{array}
\end{array}$ 
\end{small}
\\[6.pt]
\paragraph{Polynomial model}
The strain energy function of the compressible first invariant polynomial model \cite{Rivlin1951}
\begin{equation}
\psi=\sum_{i=1}^N C_{i 0}\left(\bar{I}_1-3\right)^i+\sum_{i=1}^N \frac{1}{D_i}\left(I_3-1\right)^{2 i}
\end{equation}
translates into the following parameter table
\\[6.pt]
\begin{small}
$\begin{array}{@{}l}
\mbox{{\bf{\tt{*PARAMETER TABLE, TYPE="UNIVERSAL\_TAB"}}}} \\
\begin{array}{ttttttt}
     1,&1,&1,&1,&1.0,&1.0,& C_{10} \\
     1,&1,&2,&1,&1.0,&1.0,& C_{20} \\
     1,&1,&3,&1,&1.0,&1.0,& C_{30} \\
     \vdots &\vdots &\vdots &\vdots &\vdots &\vdots & \vdots \\     
     1,&1,&N,&1,&1.0,&1.0,& C_{N0} \\
     3,&1,&2,&1,&1.0,&1.0,& 1/D_{1} \\
     3,&1,&4,&1,&1.0,&1.0,& 1/D_{2} \\
     3,&1,&6,&1,&1.0,&1.0,& 1/D_{3} \\
     \vdots &\vdots &\vdots &\vdots &\vdots &\vdots & \vdots \\  
     3,&1,&N*2,&1,&1.0,&1.0,& 1/D_{N} \\
\end{array}
\end{array}$ 
\end{small}
\\[6.pt]
\paragraph{Holzapfel model}
The strain energy function of the compressible two-fiber family Holzapfel model \cite{Holzapfel2000}
\begin{equation}
\begin{aligned}
\psi 
&= C_{10} \left( \bar{I}_1 - 3 \right) + \frac{1}{D} \left( \frac{I_3^2 - 1}{2} - \ln{I_3} \right) \\
&+ \frac{k_1}{2k_2} \left( \exp{\left[k_2 \langle \bar{I}_{4(11)}-1 \rangle^2\right]} - 1 \right) \\
&+ \frac{k_1}{2k_2} \left( \exp{\left[k_2 \langle \bar{I}_{4(22)}-1 \rangle^2\right]} - 1 \right)
\end{aligned}
\end{equation}
translates into the following six-line parameter table
\\[6.pt]
\begin{small}
$\begin{array}{@{}l}
\mbox{{\bf{\tt{*PARAMETER TABLE, TYPE="UNIVERSAL\_TAB"}}}} \\
\begin{array}{ttttttt}
     1,&1,&1,&1,&1.0,&1.0,& C_{10} \\
     4,&2,&2,&2,&1.0,&k_2,& k_1/2k_2 \\
     8,&2,&2,&2,&1.0,&k_2,& k_1/2k_2 \\
     3,&1,&1,&1,&1.0,&1.0,&1/D \\
     3,&1,&2,&1,&1.0,&0.5,&1/D \\
     3,&1,&1,&3,&1.0,&\text{-}1.0,&1/D
\end{array}
\end{array}$ 
\end{small}
\\[6.pt]
\paragraph{Kaliske model}
The strain energy function of the compressible two-fiber family Kaliske model \cite{Kaliske2004}
\begin{equation}
\begin{aligned}
\psi
&=\sum_{i=1}^3 a_i\left(\bar{I}_1-3\right)^i+\sum_{j=1}^3 b_j\left(\bar{I}_2-3\right)^j+\sum_{k=2}^6 c_k\left(\bar{I}_{4(11)}-1\right)^k \hspace*{-1.0cm}\\
&+\sum_{l=2}^6 d_l\left(\bar{I}_{5(11)}-1\right)^l +\sum_{m=2}^6 e_m\left(\bar{I}_{4(22)}-1\right)^m \\
&+\sum_{n=2}^6 f_n\left(\bar{I}_{5(22)}-1\right)^n +\frac{1}{D}\left(\frac{\left(I_3\right)^2-1}{2}-\ln (I_3)\right)
\end{aligned}
\end{equation}
translates into the following parameter table
\\[6.pt]
\begin{small}
$\begin{array}{@{}l}
\mbox{{\bf{\tt{*PARAMETER TABLE, TYPE="UNIVERSAL\_TAB"}}}} \\
\begin{array}{ttttttttt}
     1,&1,&i,&1,&1.0,&1.0,& a_i &\, &\hdots \\
     2,&1,&j,&1,&1.0,&1.0,& b_j &\, &\hdots \\
     4,&1,&k,&1,&1.0,&1.0,& c_k &\, &\hdots \\
     5,&1,&l,&1,&1.0,&1.0,& d_l &\, &\hdots \\
     8,&1,&m,&1,&1.0,&1.0,& e_m &\, &\hdots \\
     9,&1,&n,&1,&1.0,&1.0,& f_n &\, &\hdots \\
     3,&1,&1,&1,&1.0,&1.0,&1/D &&  \\
     3,&1,&2,&1,&1.0,&0.5,&1/D &&  \\
     3,&1,&1,&3,&1.0,&\text{-}1.0,&1/D && 
\end{array}
\end{array}$ 
\end{small}
\\[6.pt]
\subsection{Generalization to mixed-invariant models}
Until now, all our constitutive models have been sums of contributions of the individual invariants. It is straightforward to generalize this concept to mixed-invariant models.
Specifically, we create these mixed invariants as parameter-weighted combinations of two or more individual invariants. To incorporate mixed invariants in our material subroutine, we create a second parameter table type,
\\[6.pt]
\begin{small}
$\begin{array}{@{}l}
\mbox{{\bf{\tt{*PARAMETER TABLE TYPE, name="MIXED\_INV",}}}} \\ 
\hspace*{0.2cm} \mbox{{\bf{\tt{parameters = 16}}}} \\
\begin{array}{lll}
\mbox{\bf{\tt{INTEGER,\,,}}} &\mbox{\bf{\tt{"index n of mixed invariant,Kinv,o"}}} \\
\mbox{\bf{\tt{FLOAT,\,,}}} &\mbox{\bf{\tt{"coefficient 1st mixed invariant, K1,o"}}} \\
\mbox{\bf{\tt{FLOAT,\,,}}} &\mbox{\bf{\tt{"coefficient 2nd mixed invariant,K2,o"}}} \\
\mbox{\bf{\tt{FLOAT,\,,}}} &\mbox{\bf{\tt{"coefficient 3rd mixed invariant,K3,o"}}} \\
\hspace*{0.3cm}\vdots &\hspace*{2.0cm} \vdots  \\    
\mbox{\bf{\tt{FLOAT,\,,}}} &\mbox{\bf{\tt{"coefficient 15th mixed invariant,K15,o"}}}
\end{array} 
\end{array}$
\end{small}
\\[6.pt]
\noindent
where each of the 15 $\kappa_{j,\circ}$ mixed invariant coefficients denotes contributions to the mixed invariant $I_{\kappa,\circ}$,
\beq
\bar{I}_{\kappa,\circ}=\sum_{j=1}^{15} \kappa_{j,\circ} \, \bar{I}_j
\eeq
in which $I_j$ follows the invariant numbering in equation \ref{eq:ninv}.
To activate these mixed invariants in our FEA model, we include the following lines in our Abaqus input file,
\\[6.pt]
\begin{small}
$\begin{array}{@{}l}
\mbox{{\bf{\tt{*PARAMETER TABLE, TYPE="MIXED\_INV"}}}} \\
\begin{array}{l@{}l}
\mbox{1},&\kappa_{1,1},\kappa_{2,1},\kappa_{3,1},\kappa_{4,1},\kappa_{5,1},\kappa_{6,1},\kappa_{7,1},\kappa_{8,1},\kappa_{9,1} \\
&\kappa_{10,1},\kappa_{11,1},\kappa_{12,1},\kappa_{13,1},\kappa_{14,1},\kappa_{15,1} \\
\mbox{2},&\kappa_{1,2},\kappa_{2,2},\kappa_{3,2},\kappa_{4,2},\kappa_{5,2},\kappa_{6,2},\kappa_{7,2},\kappa_{8,2},\kappa_{9,2} \\
&\kappa_{10,2},\kappa_{11,2},\kappa_{12,2},\kappa_{13,2},\kappa_{14,2},\kappa_{15,2}\\
\hdots
\end{array}
\end{array}$ 
\end{small}
\\[6.pt]
We activate any novel constitutive neuron associated with these mixed invariants, using the following argument lines in our Abaqus input file,
\[
\begin{small}
\begin{array}{@{}l}
\mbox{{\bf{\tt{*PARAMETER TABLE, TYPE="UNIVERSAL\_TAB"}}}} \\
\begin{array}{ttttttt}
     101,&\mathrm{kf0}_{101},&\mathrm{kf1}_{101},&\mathrm{kf2}_{101},&w_{0,101},&w_{1,101},&w_{2,101} \\
     102,&\mathrm{kf0}_{102},&\mathrm{kf1}_{102},&\mathrm{kf2}_{102},&w_{0,102},&w_{1,102},&w_{2,102} \\
     \hdots
\end{array}
\end{array} 
\end{small}
\]
For clarity, we number all derived mixed invariants starting at \texttt{NINV}$=101$.

\paragraph{Holzapfel dispersion model}
The strain energy function of the Holzapfel dispersion model \cite{Gasser2006}
\begin{equation}
\begin{aligned}
\psi 
&= C_{10} \left( \bar{I}_1 - 3 \right) + \frac{1}{D} \left( \frac{I_3^2 - 1}{2} - \ln{I_3} \right) \\
&+ \frac{k_1}{2k_2} \left( \exp{\left[k_2 \langle \bar{I}_{1/4(11)}^*-1 \rangle^2\right]} - 1 \right) \\
&+ \frac{k_1}{2k_2} \left( \exp{\left[k_2 \langle \bar{I}_{1/4(22)}^*-1 \rangle^2\right]} - 1 \right)
\end{aligned}
\label{SEFGOH06}
\end{equation}
uses the two mixed invariants 
\begin{equation}
\begin{aligned}
\bar{I}_{1/4(11)}^* &= \kappa  (\bar{I}_1-3) + \left( 1 - 3\kappa \right) (\bar{I}_{4(11)}-1)\\
\bar{I}_{1/4(22)}^* &= \kappa  (\bar{I}_1-3) + \left( 1 - 3\kappa \right) (\bar{I}_{4(22)}-1)
\label{mixedinvsgoh2006}
\end{aligned}
\end{equation}
where $\kappa$ describes the dispersion of the collagen fibers ranging from $\kappa = 0.0$ for  ideally aligned fibers to $\kappa = 1/3$ for isotropically distributed fibers.
Leveraging our \mbox{\texttt{"MIXED\_INV"}} and \mbox{\texttt{"UNIVERSAL\_TAB"}} definitions, we translate this strain energy function into our universal material model subroutine by inclusion of the following  lines in our Abaqus input file,
\\[6.pt]
\begin{small}
$\begin{array}{@{}l}
\mbox{{\bf{\tt{*PARAMETER TABLE, TYPE="MIXED\_INV"}}}} \\
\mbox{{\bf{\tt{1,$\kappa$,0.0,0.0,$(1-3\kappa)$,0.0,0.0,0.0,0.0,0.0,}}}} \\
\hspace*{0.3cm} \mbox{{\bf{\tt{0.0,0.0,0.0,0.0,0.0,0.0}}}} \\
\mbox{{\bf{\tt{2,$\kappa$,0.0,0.0,0.0,0.0,0.0,0.0,$(1-3\kappa)$,0.0,}}}} \\
\hspace*{0.3cm} \mbox{{\bf{\tt{0.0,0.0,0.0,0.0,0.0,0.0}}}} \\
\mbox{{\bf{\tt{*PARAMETER TABLE, TYPE="UNIVERSAL\_TAB"}}}} \\
\begin{array}{ttttttt}
     1,&1,&1,&1,&1.0,&1.0,& C_{10} \\
     101,&2,&2,&2,&1.0,&k_2,& k_1/2k_2 \\
     102,&2,&2,&2,&1.0,&k_2,& k_1/2k_2 \\
     3,&1,&1,&1,&1.0,&1.0,&1/D \\
     3,&1,&2,&1,&1.0,&0.5,&1/D \\
     3,&1,&1,&3,&1.0,&\text{-}1.0,&1/D
\end{array}
\end{array}$
\end{small}
\\[6.pt]
\begin{figure*}[t]
\centering
\includegraphics[width=1.0\linewidth]{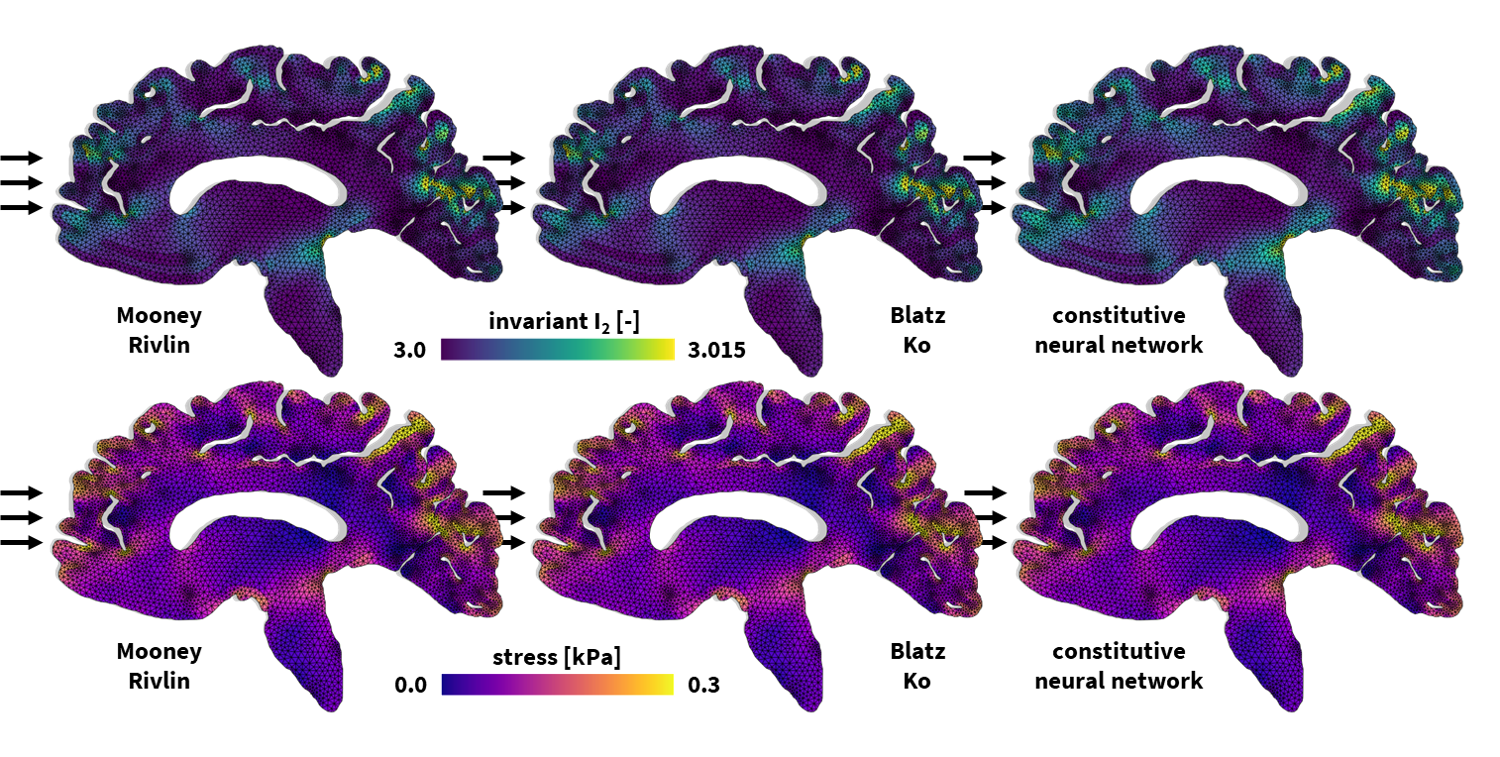}
\caption{{\textbf{\sffamily{Universal material modeling of the human brain.}} 
Deformation and stress profiles for frontal impact to the human brain. The finite element models simulate the deformation and internal tissue loading corresponding to best-fit Mooney Rivlin, Blatz Ko, and newly discovered constitutive models from left to right. All simulations leverage our universal material model subroutine and only differ in the definition of the {\tt{UNIVERSAL\_TAB}} constitutive parameter table in the finite element analysis input file.
}}\label{figbrain}
\end{figure*}
\newpage
\section{Applications}
In the following sections, we showcase examples of soft matter systems where our universal material model subroutine naturally integrates mechanical testing from the material point level to the structural level.
\subsection{The human brain}
Brain tissue is among the softest and most vulnerable tissues in the human body \cite{Budday2019}. The tissue's delicate packing of neurons, glial cells, and extracellular matrix functionally regulates most vital processes in the human body and governs human cognition, learning, and consciousness \cite{Fields2013}. As mechanics play a crucial role in neuronal function and dysfunction \cite{Goriely2015}, understanding the mechanical behavior of brain tissue is essential for anticipating how the brain will respond to injury, how it evolves during its development, or how it remodels as disease advances. Computational models play a crucial role in this endeavor, allowing researchers to simulate the multi-faceted behavior of brain tissue and explore the biomechanical role of mechanical forces in health and disease \cite{Holland2015,Lejeune2016,Noel2019,Weickenmeier2017}. These models require adequate constitutive models that capture the complex and unique characteristics of this ultrasoft, highly adaptive, and heterogeneous tissue.
\paragraph{Constitutive modeling}
Over the past decade, various research groups around the world have made significant process in the experimental and constitutive characterization of human brain tissue \cite{Budday2017}. This has led to multiple competing constitutive models to characterize the behavior of gray and white matter tissue. Most notably, neo Hooke \cite{Treloar1948}, Blatz Ko \cite{Blatz1962}, Mooney Rivlin \cite{Mooney1940,Rivlin1948}, Demiray \cite{Demiray1972}, Gent \cite{Gent1996}, and Holzapfel \cite{Holzapfel2000} models were proposed as successful candidates to characterize the stress-stretch response of these tissues. Given brain tissue's intricate behavior, fitting a constitutive model to one single loading mode, tension, compression, or shear, does not generalize well to the other modes \cite{Linka2023b,StPierre2023}. Therefore, we consider a widely-used benchmark dataset where $5 \times 5 \times 5$mm$^3$ human brain samples were tested in tension, compression, and shear \cite{Budday2017,Budday2017a,Budday2019}. We concomitantly discover and fit the best possible constitutive models considering these loading modes together and find the following three best models and parameters \cite{Linka2023b}. \\[6.pt]
\noindent
The Mooney Rivlin model \cite{Mooney1940,Rivlin1948}
\[
\psi=\frac{1}{2}\mu_1\left(\bar{I}_1-3\right)+\frac{1}{2}\mu_2\left(\bar{I}_2-3\right)
\]
with parameters $\mu_1=0.0021$ kPa, $\mu_2=1.8817$ kPa for the gray matter cortex, and $\mu_1=0.0168$ kPa, $\mu_2=0.9697$ kPa for the white matter corona radiata. This translates into
\\[6.pt]
\begin{small}
$\begin{array}{@{}l}
\mbox{{\bf{\tt{*PARAMETER TABLE, TYPE="UNIVERSAL\_TAB"}}}} \\
\begin{array}{ttttttt}
    1,&1,&1,&1,&1.0,&1.0,& \mu_1/2 \\
    2,&1,&1,&1,&1.0,&1.0,& \mu_2/2 
\end{array}
\end{array}$ 
\end{small}
\\[12.pt]
\noindent
The Blatz Ko model \cite{Blatz1962}
\[
\psi=\frac{1}{2}\mu\left(\bar{I}_2-3\right)
\]
with parameters $\mu=1.9043$ kPa for the gray matter cortex, and $\mu=0.9556$ kPa for the white matter corona radiata. This translates into
\\[6.pt]
\begin{small}
$\begin{array}{@{}l}
\mbox{{\bf{\tt{*PARAMETER TABLE, TYPE="UNIVERSAL\_TAB"}}}} \\
\mbox{{\bf{\tt{2,1,1,1,1.0,1.0,$\mu$/2 }}}} \\ 
\end{array}$ 
\end{small}
\\[12.pt]
\noindent
Our newly discovered six-term model \cite{Linka2023b,Peirlinck2024}
\[
\begin{aligned}
\psi 
&=  \mu_1 \left[\bar{I}_2-3\right] +\frac{a_1}{2b_1}\left[\exp \left(b_1\left[\bar{I}_2-3\right]\right)-1\right] \\
&-\frac{\alpha_1}{2\beta_1} \ln \left(1-\beta_1\left[\bar{I}_2-3\right]\right) +\mu_2\left[\bar{I}_2-3\right]^2 \\
& +\frac{a_2}{2b_2}\left[\exp \left(b_2\left[\bar{I}_2-3\right]^2\right)-1\right] 
-\frac{\alpha_2}{2\beta_2} \ln \left(1-\beta_2\left[\bar{I}_2-3\right]^2\right)
\end{aligned}
\]
with non-zero terms
$\alpha_1=1.2520$ kPa,
$\beta_1=0.9875$,
$\mu_2=3.8007$ kPa,
$a_2=6.2285$ kPa,
$b_2=1.6495$,
$\alpha_2=4.6743$ kPa, and
$\beta_2=1.6663$ 
for the gray matter cortex and
$\mu_1=0.2215$ kPa,
$a_1=0.2350$ kPa,
$b_1=0.2398$,
$a_2=6.3703$ kPa,
$b_2=1.8893$,
$\alpha_2=4.5065$ kPa, and
$\beta_2=1.1789$ for the white matter corona radiata. 
We translate this model into the following six-line parameter table of our universal material model:
\\[6.pt]
\begin{small}
$\begin{array}{@{}l}
\mbox{{\bf{\tt{*PARAMETER TABLE, TYPE="UNIVERSAL\_TAB"}}}} \\
\begin{array}{ttttttt}
    2,&1,&1,&1,&1.0,&1.0,& \mu_1 \\
    2,&1,&1,&2,&1.0,&b_1,& a_1/2b_1 \\
    2,&1,&1,&3,&1.0,&\beta_1,& \alpha_1/2\beta_1 \\
    2,&1,&2,&1,&1.0,&1.0,& \mu_2 \\
    2,&1,&2,&2,&1.0,&b_2,& a_2/2b_2 \\
    2,&1,&2,&3,&1.0,&\beta_2,& \alpha_2/2\beta_2 \\
\end{array}
\end{array}$ 
\end{small}
\\[6.pt]
\paragraph{Simulation}
Utilizing our universal material model subroutine, we incorporate these brain models into a realistic vertical head impact finite element simulation \cite{Peirlinck2024}. Based on magnetic resonance images \cite{Harris2018}, 
we create the two-dimensional sagittal finite element model in Figure \ref{figbrain}. In this model, gray and white matter are spatially discretized using 6,182 gray and 5,701 white linear triangular elements, resulting in 6,441 nodes, and 12,882 degrees of freedom in total. We embed our model into the skull using spring support at the free boundaries and apply a frontal impact to the brain that we represent with all three models, the Mooney Rivlin, Blatz Ko, and new discovered models, as shown in Figure \ref{figbrain}.
\subsection{Skin}
Skin is the largest organ of the human body \cite{Tepole2011}. 
It serves vital functions for our survival such as being the first line of defense against mechanical injury while at the same time allowing us to move and interact with the world \cite{Limbert2014}. 
Surgery of any kind entails skin rupture and manipulation \cite{Aarabi2007}. 
Especially during reconstructive procedures, skin tissues are subjected to extreme deformations \cite{Lee2020}. 
The complex stress field generated by skin tissue manipulation has a direct effect on the subsequent wound healing response, with excessive stress causing increased inflammatory response that can lead to fibrosis \cite{Lee2018}. 
In some cases, excessive stress can even result in tissue necrosis \cite{Gosain2018}. 
Thus, accurate computational models of skin are key to design safe reconstructive surgical procedures. 
\begin{figure*}[t]
\centering
\includegraphics[width=1.0\linewidth]{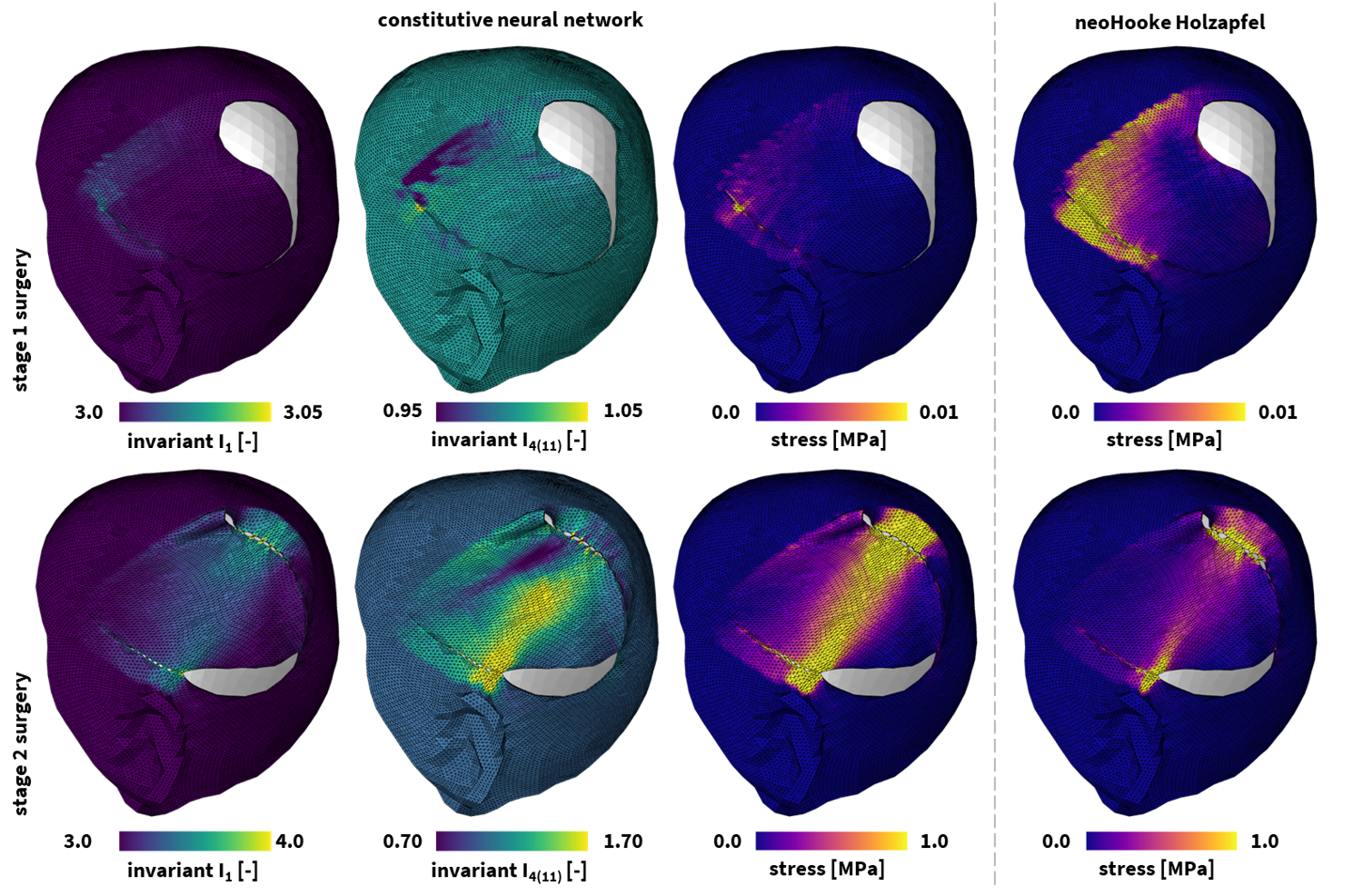}
\caption{{\textbf{\sffamily{Universal material modeling of skin.}} 
Deformation and stress profiles in the human scalp following a melanoma resection reconstruction procedure. The finite element models simulate the deformation and internal tissue loading corresponding a two-stage flap rotation and suturing procedure, with the first stage shown in the top row and the second stage shown in the bottom row. The remaining wound is closed with a skin graft to avoid excessive tissue stresses and damage. Both tissue manipulations are modeled using the best-fit constitutive neural network model in the three left columns. For comparison, we also showcase the resulting stress profiles for the best-fit neo Hooke Holzapfel model in the right column. All simulations leverage our universal material model subroutine and only differ in the definition of the {\tt{UNIVERSAL\_TAB}} constitutive parameter table in the finite element analysis input file.
}}\label{figskin}
\end{figure*}
\paragraph{Constitutive modeling}
Skin modeling has received significant attention for more than half a century \cite{Lanir1976,Tong1976}. 
Isotropic models such as the neo Hooke \cite{Treloar1948} or Mooney Rivlin \cite{Mooney1940,Rivlin1948} models have been used, but show significant limitations. Not only do they fail to describe the anisotropy of skin, they also lack the ability to capture this tissue's rapid strain-stiffening behavior \cite{Lanir1976}.
To overcome these issues, we examine combined uniaxial and biaxial tensile testing data of porcine skin tissue samples \cite{Tac2022,Tac2022a} to discover more accurate material models that depict the anisotropic stress-stretch behavior. First, we fit the microstructure-inspired Holzapfel model \cite{Holzapfel2000},  
\[
    \psi=\frac{1}{2} \mu\left[\bar{I}_1-3\right]+\frac{1}{2} \frac{a_4}{b_4}\left[\exp \left(b_4\langle \bar{I}_{4(11)}-1\rangle^2\right)-1\right]\,.
\]
This model was originally developed for arterial tissues and combines the isotropic linear first invariant neo Hooke term, $\left[\bar{I}_1-3\right]$, with an anisotropic quadratic exponential fourth invariant term, $\langle \bar{I}_{4(11)}-1\rangle$, along the collagen fiber direction. Here, our best possible fit to the combined uniaxial and biaxial testing data results in
$\mu = 0.2492$ MPa,
$a_4 = 0.1054$ MPa, and
$b_4 = 10.7914$.
We naturally incorporate this constitutive model and parameters in our universal material model subroutine using the following two-line parameter table
\\[6.pt]
\begin{small}
$\begin{array}{@{}l}
\mbox{{\bf{\tt{*PARAMETER TABLE, TYPE="UNIVERSAL\_TAB"}}}} \\
\begin{array}{ttttttt}
    1,&1,&1,&1,&1.0,&1.0,& \mu \\
    4,&2,&2,&2,&1.0,&b_4,& a_4/2b_4
\end{array}
\end{array}$ 
\end{small}
\\[6.pt]
Given the rather low mean goodness of fit $\rm{R}^2 = 0.6857$, we subsequently leverage a tranversely isotropic constitutitive neural network to discover a more accurate model \cite{Linka2023a}.
From a library of $2^{16}=65,536$ possible combinations of terms, we discover a model in two exponential quadratic terms,
\[
\begin{aligned}
\psi 
& = \frac{a_1}{2b_1} \left( \exp{\left[b_1 \left(\bar{I}_{1}-3 \right)^2 \right]} - 1 \right) \\
& + \frac{a_4}{2b_4} \left( \exp{\left[b_4 \langle \bar{I}_{4(11)}-1 \rangle^2 \right]} - 1 \right)
\label{SEFCANNskin}
\end{aligned}
\]
with parameters $a_1 = 1.3291$ MPa, $b_1 =0.8207$, $a_4= 0.2656$ MPa, and $b_4= 0.3921$ \cite{Linka2023a}. 
To integrate this new model into a finite element simulation, we incorporate the following two parameter lines in our universal material subroutine 
\\[6.pt]
\begin{small}
$\begin{array}{@{}l}
\mbox{{\bf{\tt{*PARAMETER TABLE, TYPE="UNIVERSAL\_TAB"}}}} \\
\begin{array}{ttttttt}
    1,&1,&2,&2,&1.0,&b_1,& a_1/2b_1 \\
    4,&2,&2,&2,&1.0,&b_4,& a_4/2b_4
\end{array}
\end{array}$ 
\end{small}
\\[6.pt]
In contrast to the neo Hooke Holzapfel model, this discovered constitutive neural network model has a mean good of fit $\rm{R}^2 = 0.8629$ for the combined uniaxial and biaxial porcine skin testing data \cite{Linka2023a}.
\paragraph{Simulation}
Leveraging our universal material subroutine, we integrate both material models in a finite element simulation of a 62-year-old adult male patient undergoing reconstructive surgery following surgical melanoma resection \cite{Lee2020}. 
A three-dimensional patient specific geometry was obtained via multi-view stereo reconstruction of a sequence of photos taken in the operating room before and after surgery. 
The scalp was approximated based on the skin surface and spatially discretized using 75,282 linear tetrahedral elements and 25,394 nodes, leading to a total 76,182 degrees of freedom. 
Our simulation recapitulates the closure of the resected tissue defect by imposing nodal constraints to nodes on either edge of the defect to mimic sutures used to close the wound. 
Figure \ref{figskin} showcases the deformation and internal tissue tension profiles following the two-step surgical skin reconstruction procedure. 
We clearly observe the limited tissue deformation and loading profiles during the first stage in the top row. In contrast, during the second stage surgery in the bottom row, substantial deformations develop across the skin. 
Specifically, we appreciate the regional differences between the isotropic $\bar{I}_1$ and anisotropic $\bar{I}_{4(11)}$ deformation invariants. 
Figure \ref{figskin} also showcases noticeable stress profile differences between the newly discovered material model and the neo Hooke Holzapfel model in the third and fourth columns.
In the lower stretch regimes shown in the first stage reconstruction, the neo Hooke Holzapfel model clearly overestimates the stresses in the skin. In the higher stretch regimes, shown during the second stage reconstruction in the bottom, the neo Hooke Holzapfel fit underestimates the stresses in the tissue. 
While a modeling-based overestimation of the stress state holds limited risks from a medical point of view, an underestimation could have harmful consequences as clinical decisions co-informed by such models could cause excessive tissue damage and scarring. 
Figure \ref{figskin} showcases the crucial aspect that proper constitutive modeling and calibration plays in this regard, in which the neo Hooke Holzapfel model, which does not properly capture skin tissue's strain-stiffening, underestimates the tissue stress in comparison to the more accurate newly discovered model.
\subsection{Human arteries}
\begin{figure*}[t]
\centering
\includegraphics[width=1.0\linewidth]{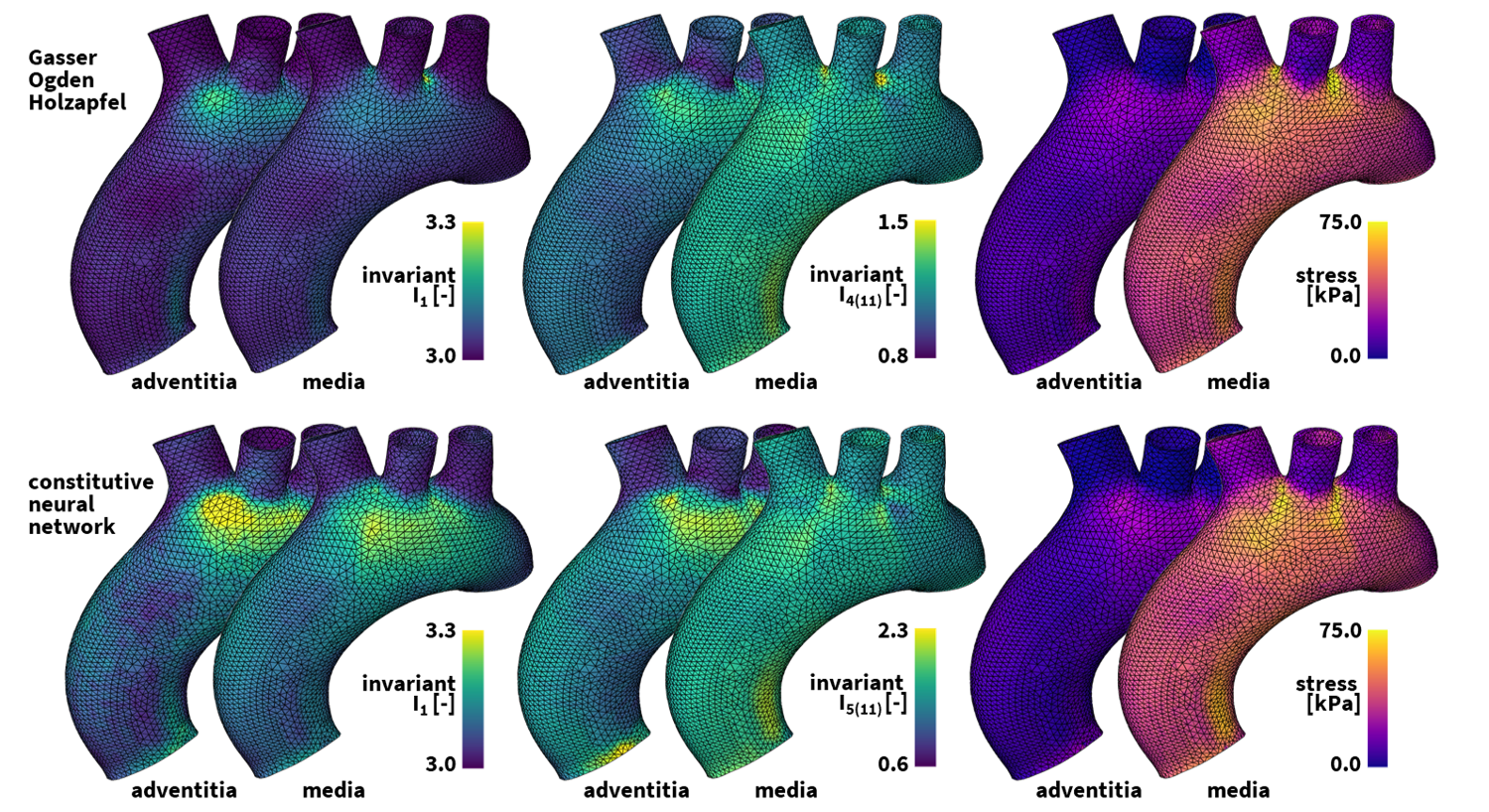}
\caption{{\textbf{\sffamily{Universal material modeling of human arteries.}} 
Diastolic deformation and stress profiles in the media and adventitia layer of the human ascending aortic arch. The finite element models simulate the deformation and internal tissue loading corresponding to the best-fit Holzapfel dispersion model in the top row and newly discovered model in the bottom row. Both simulations leverage our universal material model subroutine and only differ in the definition of the {\tt{UNIVERSAL\_TAB}} constitutive parameter table in the finite element analysis input file.
}}\label{figarterial}
\end{figure*}
Computational simulations play a pivotal role in understanding and predicting the biomechanical factors of a wide variety of arterial diseases \cite{Humphrey2021,Peirlinck2019,Updegrove2016}.
In vascular medicine, knowing the precise stress and strain fields across the vascular wall is critical for understanding the formation, growth, and rupture of aneurysms and dissections \cite{Humphrey2012,VanderLinden2023,Gheysen2024}; 
for identifying high-risk regions of plaque formation, rupture, and thrombosis \cite{Holzapfel2005,Rausch2017}; 
and for optimizing stent design and surgery \cite{DeBock2012,Famaey2012}.
\paragraph{Constitutive modeling}
Over the past four decades, various phenomenological polynomial \cite{Vaishnav1973,vonMaltzahn1981}, exponential \cite{Fung1979}, logarithmic \cite{Takamizawa1987},  and exponential-polynomial \cite{Kasyanov1980,Peirlinck2023,Zhou1997} models have been proposed to describe the non-linear elastic, anisotropic, quasi-incompressible behavior of arterial tissue.
Recently, microstructurally-informed models were brought forward, including symmetric two- and four-fiber family models \cite{Gasser2006,Holzapfel2000,Baek2007}, either symmetric or unsymmetric \cite{Holzapfel2015}. 
All these material models can fit uniaxial and biaxial arterial tissue testing data, but do not always generalize well to off-axis testing regimes \cite{Schroeder2018}. \\[6.pt]
We consider biaxial tensile testing of thoracic aortic tissue samples at five differing circumferential-axial stretch ratios \cite{Niestrawska2016, Niestrawska2018}
Using data-driven constitutive neural networks, we discover the most appropriate arterial material model. From a library of $2^{16}=65,536$ possible combinations of terms, we discover 
\[
\begin{aligned}
  \psi 
&= \frac{\mu_1}{2} \left[\bar{I}_{1}-3\right] 
+ \frac{a}{2b} \left( \exp{\left[b \left(\bar{I}_{1}-3 \right) \right]} - 1 \right) \\
&+ \sum_{i=1,2} \frac{1}{2} \mu_5\left\langle \bar{I}_{5(i i)}-1\right\rangle^2
\end{aligned}
\]
with an isotropic linear and exponential linear first invariant term and an anisotropic quadratic fifth invariant term. 
Our best-fit parameters read 
$\mu_1 = 33.45$ kPa, $a = 3.74$ kPa, $b = 6.66$, $\mu_5 = 2.17$ kPa for the media at an angle $\alpha = \pm7.00^{\circ}$ and
$\mu_1 =  8.30$ kPa, $a = 1.42$ kPa, $b = 6.34$, $\mu_5 = 0.49$ kPa for the adventitia at an angle $\alpha = \pm66.78^{\circ}$.
This translates into the following four-line parameter table of our universal material model,
\\[6.pt]
\begin{small}
$\begin{array}{@{}l}
\mbox{{\bf{\tt{*PARAMETER TABLE, TYPE="UNIVERSAL\_TAB"}}}} \\
\begin{array}{ttttttt}
    1,&1,&1,&1,&1.0,&1.0,& \mu_1/2 \\
    1,&1,&1,&2,&1.0,&b,& a/2b \\
    5,&2,&2,&1,&1.0,&1.0,& \mu_5/2 \\
    9,&2,&2,&1,&1.0,&1.0,& \mu_5/2 \\
\end{array}
\end{array}$ 
\end{small}
\\[6.pt]
Alternatively, in the classical microstructure-inspired dispersion type Holzapfel model \cite{Holzapfel2005}
\[
\psi=\frac{1}{2} \mu\left[\bar{I}_1-3\right]+\sum_{i=1,2} \frac{a}{2b} \left( \exp{\left[b \langle \bar{I}_{1/4(ii)}^*-1 \rangle^2\right]} - 1 \right)
\]
our best-fit parameters are 
$\mu = 48.68$ kPa, $a = 6.67$ kPa, $b = 23.17$, $\kappa = 0.074$ for the media at $\alpha = \pm7.00^{\circ}$) and
$\mu = 13.22$ kPa, $a = 0.93$ kPa, $b = 12.06$, $\kappa = 0.091$ for the adventitia at $\alpha = \pm66.78^{\circ}$.
We translate this model into the following parameter table of our universal material model 
\\[6.pt]
\begin{small}
$\begin{array}{@{}l}
\mbox{{\bf{\tt{*PARAMETER TABLE, TYPE="MIXED\_INV"}}}} \\
\mbox{{\bf{\tt{1,$\kappa$,0.0,0.0,$(1-3\kappa)$,0.0,0.0,0.0,0.0,0.0,}}}} \\
\hspace*{0.3cm} \mbox{{\bf{\tt{0.0,0.0,0.0,0.0,0.0,0.0}}}} \\
\mbox{{\bf{\tt{1,$\kappa$,0.0,0.0,0.0,0.0,0.0,0.0,$(1-3\kappa)$,0.0,}}}} \\
\hspace*{0.3cm} \mbox{{\bf{\tt{0.0,0.0,0.0,0.0,0.0,0.0}}}} \\
\mbox{{\bf{\tt{*PARAMETER TABLE, TYPE="UNIVERSAL\_TAB"}}}} \\
\begin{array}{ttttttt}
     1,&1,&1,&1,&1.0,&1.0,& \mu/2 \\
     101,&2,&2,&2,&1.0,&b,& a/2b \\
     102,&2,&2,&2,&1.0,&b,& a/2b
\end{array}
\end{array}$ 
\end{small}
\\[6.pt]
\paragraph{Simulation}
Using our universal material subroutine, we integrate both models in a finite element simulation of the human aortic arch under hemodynamic loading conditions \cite{Peirlinck2018}. 
Our aortic arch geometry is extracted from high-resolution magnetic resonance images of a healthy, 50th percentile U.S. male \cite{Peirlinck2021}. 
We assume an average aortic wall thickness of 3.0 mm, where the inner 75\% of the wall make up the media and the outer 25\% make up the adventitia. 
We discretize our geometry using 60,684 linear tetrahedral elements for the media and 30,342 linear tetrahedral elements for the adventitia, leading to a total 61,692 degrees of freedom. 
The local collagen fiber angles against the circumferential direction are $\pm$ 7.00$^{\circ}$ in the media and $\pm$ 66.78$^{\circ}$ in the adventitia and are locally defined as a vector field variable for each element. 
We use continuum distributed coupling boundary conditions at the aortic outlets to constrain the arch in space \cite{Peirlinck2018a}, and leverage Neumann boundary conditions to simulate the hemodynamic loading conditions the aortic arch undergoes during a single cardiac cycle. 
Figure \ref{figarterial} showcases the computed diastolic stresses in the media and the adventitia for both our newly discovered model and the microstructure-informed dispersion-type Holzapfel model \cite{Peirlinck2023}. 
\subsection{Heart valves}
\begin{figure*}[h]
\centering
\includegraphics[width=1.0\linewidth]{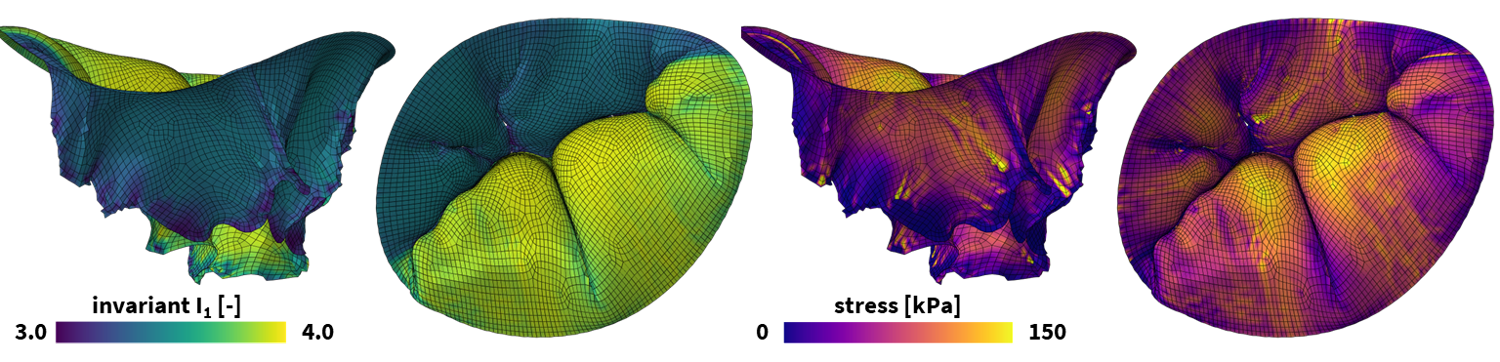}
\caption{{\textbf{\sffamily{Universal material modeling of heart valves.}} 
Personalized tricuspid valve loading during the cardiac cycle. The finite element models simulate the deformation, left, and internal tissue loading, right, in response to the inter-ventricular pressure changes from end-diastole to end-systole. The tricuspid valve is shown from a side and top view. Each valvular leaflet leverages our universal material model subroutine and only differs in the definition of the {\tt{UNIVERSAL\_TAB}} constitutive parameter table in the finite element analysis input file.
}}\label{figvalve}
\end{figure*}
The tricuspid valve is our right atrioventricular valve which ensures unidirectional blood flow through the right side of the heart. 
Often as a result of other primary diseases \cite{Rogers2010,Dreyfus2015}, a diseased tricuspid valve can fail to close and regurgitate. 
Tricuspid valve disease affects over one million Americans and is associated with increased patient mortality and morbidity \cite{Nath2004,Nkomo2006}. 
Computational models of the tricuspid valve provide valuable insights into the workings of the valve, and have been used to increase our understanding of the progression of valve disease \cite{Mathur2023} and to work towards improved repair outcomes \cite{Haese2023}.
\paragraph{Constitutive modeling}
Numerous studies have investigated the mechanical behavior of atrioventricular valve leaflets. Valvular leaflets exhibit a pronounced anisotropy and a non-linear behavior, motivating an anisotropic exponential material model to capture this complex material behavior \cite{Lee2014}. 
Others have used microstructurally-informed models \cite{Gasser2006,Kong2018} or anisotropic exponential Fung-type models \cite{Khoiy2018} to capture the material response of the tricuspid valve leaflets. 
However, the tricuspid valve leaflets only exhibit slight anisotropy \cite{Pham2017}. 
To improve the ease of use in computational models, recent studies have proposed a 
simplified isotropic Fung-type exponential function \cite{Kamensky2021}. 
Leveraging force-controlled 400 mN equibiaxial mechanical tests on 7$\times$7 mm valve leaflet tissue samples \cite{Meador2020}, we fit the following two-term isotropic exponential Fung-Type model \cite{Mathur2022}
\[
\psi = \, \frac{c_0}{2} \left[\bar{I}_{1}-3\right] + \frac{c_1}{2} \left( \exp{\left[c_2 \left(\bar{I}_{1}-3 \right)^2 \right]} - 1 \right)
\]
with an isotropic linear first invariant term describing the response at small-strains and under compression and an exponential first invariant term determining the strain-stiffening response under large strains \cite{Kamensky2021}.
Our best-fit parameters are
$c_0=1.0$ kPa, $c_1=0.124$ kPa, $c_2= 4.57$ for the anterior, 
$c_0=1.0$ kPa, $c_1=0.188$ kPa, $c_2=14.86$ for the posterior, and
$c_0=1.0$ kPa, $c_1=0.191$ kPa, $c_2=17.75$ for the septal leaflets.
To incorporate this constitutive model in our universal material subroutine, we define the following two parameter lines
\\[6.pt]
\begin{small}
$\begin{array}{@{}l}
\mbox{{\bf{\tt{*PARAMETER TABLE, TYPE="UNIVERSAL\_TAB"}}}} \\
\begin{array}{ttttttt}
     1,&1,&1,&1,&1.0,&1.0,& c_0/2 \\
     1,&1,&2,&2,&1.0,&c_2,& c_1/2
\end{array}
\end{array}$ 
\end{small}
\\[6.pt]
\paragraph{Simulation}
Using our universal material subroutine, we integrate the constitutive behavior of all three leaflets into a personalized finite element model of the tricuspid valve, the Texas 1.1 TriValve \cite{Mathur2022,TriValve2024}.
Through personalized pressure and annular displacement recordings in the realistic hemodynamic environment of an organ preservation system and image-based planimetry on the excised valve, a three-dimensional reconstruction of the tricuspid valve is build at end-diastole. 
The valve and chordae geometries are spatially discretized using 8,283 linear quadrilaterial shell elements and 4,169 three-dimensional linear multi-segmented truss elements, resulting in a total 25,761 degrees of freedom. 
By imposing the recorded personalized annular displacements and an end-systolic transvalvular pressure of 22.95 mmHg on the ventricular surface of the valve, we simulate valvular loading from end-diastole to end-systole. 
Figure \ref{figvalve} showcases the resulting deformation and maximum principal stress contours the tricuspid valve. Notably, the varying stiffnesses of the anterior, septal, and posterior leaflets result in noticeable differences in the first invariant of the Cauchy-Green deformation tensor, but in comparable maximum principal stress profiles across the leaflets.
\subsection{The human heart}
\begin{figure*}[h]
\centering
\includegraphics[width=1.0\linewidth]{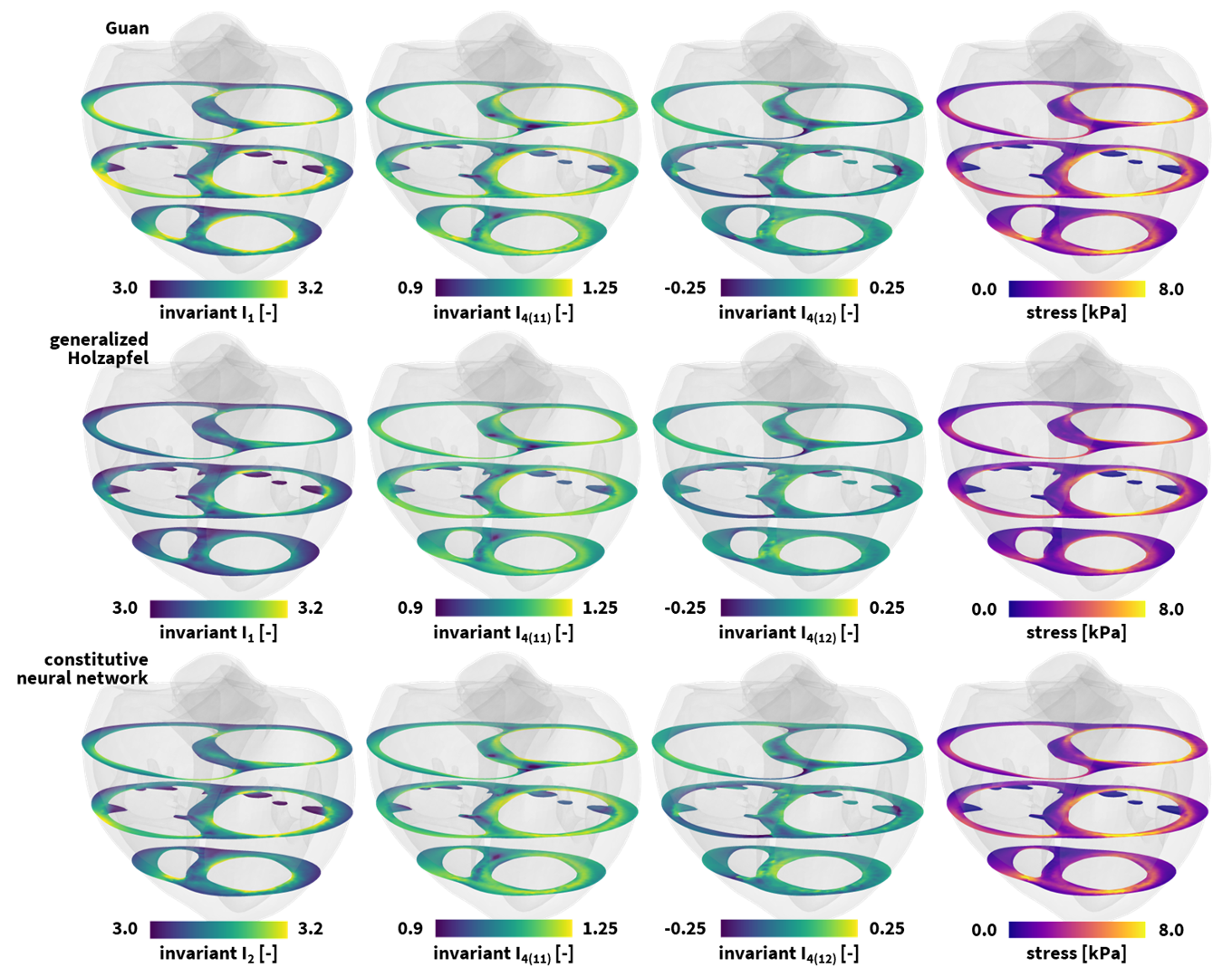}
\caption{{\textbf{\sffamily{Universal material modeling of the human heart.}} 
Personalized isotropic and directional deformation invariant and maximum principal stresses stress profiles, in short-axis slices frontal views, resulting from a healthy left and right ventricular end-diastolic pressure loading of 8mmHg and 4mmHg. The finite element models simulate the deformation and internal tissue loading corresponding to the best-fit Guan model in the top row, the generalized Holzapfel model in the middle row, and the newly discovered model in the bottom row. All three simulations leverage our universal material model subroutine and only differ in the definition of the {\tt{UNIVERSAL\_TAB}} constitutive parameter table in the finite element analysis input file.
}}
\label{figheart}
\end{figure*}
Cardiac disorders are a leading cause of morbidity and death worldwide \cite{Mensah2019}. Computational models of cardiac function hold immense potential to contribute to our understanding of health and disease, improve our diagnostic analyses, and optimize personalized intervention \cite{Peirlinck2021,Augustin2021,Fedele2023}. 
For example, corrective surgeries in obstructive cardiomyopathy \cite{Quarteroni2023} and congenital heart defects \cite{Tikenogullari2023}, the replacement of diseased valves \cite{Fumagalli2023}, or the implantation of a cardiac assist device \cite{Sack2018} all involve complex and delicate procedures that demand careful planning and simulation to ensure their success. 
Crucially, the accuracy and reliability of these computational models hinge on precise constitutive modeling of the underlying mechanical behavior of myocardial tissue. 
\paragraph{Constitutive modeling}
Research on constitutive models that accurately describe passive myocardial mechanics spans over five decades. 
One of the earliest models described cardiac muscle tissue as an isotropic hyperelastic material \cite{Demiray1976}. 
Later, with increasing experimental insights, more sophisticated transversely isotropic \citep{Humphrey1987, Costa1996}, and eventually orthotropic \citep{Schmid2007, Holzapfel2009} constitutive models were introduced. 
This latter cardiac-tissue Holzapfel model is currently one of the most popular models for heart muscle tissue and fits simple shear tests of myocardial tissue well \cite{Dokos2002}. 
Nevertheless, it displays limitations when simultaneously fitted to different loading modes \cite{Guan2019}. 
Therefore, we consider triaxial shear and biaxial extension tests on human myocardial tissue \cite{Sommer2015}, and use these data to discover and the best possible model and parameters to characterize both loading conditions combined \cite{Martonova2024}.\\[6.pt]
We begin with the four-term Guan model \cite{Guan2019} 
that features 
an exponential linear term in the first invariant $\bar{I}_1$,
exponential quadratic terms in the fiber and normal fourth invariants $\bar{I}_{\rm{4f}}$ and $\bar{I}_{\rm{4n}}$,
and an exponential quadratic term in the fiber-sheet eighth invariant $\bar{I}_{\rm{8fs}}$,
\[
\begin{aligned}
\psi &= \frac{a}{2b}\left[ \,\exp \left(b[\bar{I}_1 - 3 ]\right) \right] +\frac{a_{\rm{f}}}{2b_{\rm{f}}}\,\left[ \, \exp \left(b_{\rm{f}} \langle \bar{I}_{\rm{4f}} -1 \rangle^2 \right) - 1 \right] \\
&+ \frac{a_{\rm{n}}}{2b_{\rm{n}}}\,[ \, \exp \left(b_{\rm{n}} \langle \bar{I}_{\rm{4n}} -1 \rangle^2 \right) - 1]+\frac{a_{\rm{fs}}}{2b_{\rm{fs}}} [ \, \exp \left( b_{\rm{fs}} [ I_{\rm{8fs}}  ]^2\right) - 1].  
\end{aligned}
\]
Calibrating this model simultaneously on biaxial tensile and triaxial shear data for human myocardial tissue, we obtain
a mean goodness of fit $\rm{R}^2 = 0.867$
for parameters
$a = 0.782$ kPa,
$b = 7.248$,
$a_{\rm{f}} = 4.488$ kPa,
$b_{\rm{f}} = 14.571$,
$a_{\rm{n}} = 2.513$ kPa,
$b_{\rm{n}} = 10.929$,
$a_{\rm{fs}} = 0.436$ kPa, and
$b_{\rm{fs}} = 4.959$.
To incorporate this constitutive model in our universal material subroutine, we define the following four parameter lines,
\begin{flushleft}
\begin{small}
$\begin{array}{@{}l}
\mbox{{\bf{\tt{*PARAMETER TABLE, TYPE="UNIVERSAL\_TAB"}}}} \\
\begin{array}{ttttttt}
     1,&1,&1,&2,&1.0,&b,& a/2b \\
     4,&2,&2,&2,&1.0,&b_f,& a_f/2b_f \\
     14,&2,&2,&2,&1.0,&b_n,& a_n/2b_n \\
     6,&1,&2,&2,&1.0,&b_{fs},& a_{fs}/2b_{fs}
\end{array}
\end{array}$ 
\end{small}
\end{flushleft}
Next, we consider the seven-term generalized Holzapfel model \cite{Holzapfel2009} which features an exponential linear term in the first invariant $\bar{I}_1$,
exponential quadratic terms of all fourth invariants $\bar{I}_{\rm{4f}}$, $\bar{I}_{\rm{4s}}$, $\bar{I}_{\rm{4n}}$,
and an exponential quadratic term in all eighth invariants $\bar{I}_{\rm{8fs}}$, $\bar{I}_{\rm{8fn}}$, $\bar{I}_{\rm{8sn}}$. 
\[
\begin{aligned}
\psi &= \frac{a}{2b}\left[ \,\exp \left(b[\bar{I}_1 - 3 ]\right) \right] +\frac{a_{\rm{f}}}{2b_{\rm{f}}}\,\left[ \, \exp \left(b_{\rm{f}} \langle \bar{I}_{\rm{4f}} -1 \rangle^2 \right) - 1 \right] \\
&+\frac{a_{\rm{s}}}{2b_{\rm{s}}}\,[ \, \exp \left(b_{\rm{s}} \langle \bar{I}_{\rm{4s}} -1 \rangle ^2 \right) - 1] + \frac{a_{\rm{n}}}{2b_{\rm{n}}}\,[ \, \exp \left(b_{\rm{n}} \langle \bar{I}_{\rm{4n}} -1 \rangle^2 \right) - 1]\\
&+\frac{a_{\rm{fs}}}{2b_{\rm{fs}}} [ \, \exp \left( b_{\rm{fs}} [\bar{I}_{\rm{8fs}}  ]^2\right) - 1] + \frac{a_{\rm{sn}}}{2b_{\rm{sn}}} [ \, \exp \left(b_{\rm{sn}} [ \bar{I}_{\rm{8sn}}  ]^2 \right) - 1].  
\end{aligned}
\]
A combined triaxial-biaxial training of this model 
calibrates the model parameters to
$a = 0.950$ kPa,
$b = 5.457$,
$a_{\rm{f}} = 3.318$ kPa,
$b_{\rm{f}} = 23.701$,
$a_{\rm{s}} = 1.405$ kPa,
$b_{\rm{s}} = 20.067$,
$a_{\rm{n}} = 2.037$ kPa,
$b_{\rm{n}} = 16.976$,
$a_{\rm{fs}} = 0.586$ kPa,
$b_{\rm{fs}} = 1.081$,
$a_{\rm{sn}} = 0.047$ kPa, and
$b_{\rm{sn}} = 11.842$. 
This model has a mean goodness of fit $\rm{R}^2 = 0.876$ \cite{Martonova2024}.
We translate this constitutive model into our universal material subroutine through the definition of the following parameter lines in our finite element analysis input  file
\begin{flushleft}
\begin{small}
$\begin{array}{@{}l}
\mbox{{\bf{\tt{*PARAMETER TABLE, TYPE="UNIVERSAL\_TAB"}}}} \\
\begin{array}{ttttttt}
     1,&1,&1,&2,&1.0,&b,& a/2b \\
     4,&2,&2,&2,&1.0,&b_f,& a_f/2b_f \\
     8,&2,&2,&2,&1.0,&b_s,& a_s/2b_s \\
     14,&2,&2,&2,&1.0,&b_n,& a_n/2b_n \\
     6,&1,&2,&2,&1.0,&b_{fs},& a_{fs}/2b_{fs} \\
     12,&1,&2,&2,&1.0,&b_{sn},& a_{sn}/2b_{sn} \\
\end{array}
\end{array}$ 
\end{small}
\end{flushleft}
Finally, we leverage an orthotropic constitutive neural network to discover the best model and parameters to explain the experimental data.
From a library of $2^{32}=4,294,967,296$ possible combinations of terms and a sparsity-promoting regularization with $\alpha=0.01$, we discover a four-term model,
\[
\begin{aligned}
\psi &= \mu \left(\bar{I}_2 - 3\right)^2 +\frac{a_{\rm{f}}}{2b_{\rm{f}}}\,\left[ \, \exp \left(b_{\rm{f}} \langle \bar{I}_{\rm{4f}} -1 \rangle^2 \right) - 1 \right] \\
&+ \frac{a_{\rm{n}}}{2b_{\rm{n}}}\,[ \, \exp \left(b_{\rm{n}} \langle \bar{I}_{\rm{4n}} -1 \rangle^2 \right) - 1]+\frac{a_{\rm{fs}}}{2b_{\rm{fs}}} [ \, \exp \left( b_{\rm{fs}} [ \bar{I}_{\rm{8fs}}  ]^2\right) - 1].  
\end{aligned}
\]
with a mean goodness of fit $\rm{R}^2 = 0.894$ \cite{Martonova2024}.
Here, our discovered material parameters amount to
$\mu = 5.162$ kPa,
$a_{\rm{f}} = 3.426$ kPa,
$b_{\rm{f}} = 21.151$,
$a_{\rm{n}} = 2.754$ kPa,
$b_{\rm{n}} = 4.371$,
$a_{\rm{fs}} = 0.494$ kPa, and
$b_{\rm{fs}} = 0.508$.
We integrate this newly discovered model for myocardial tissue in our finite element analysis  through the following four-line parameter table 
\begin{flushleft}
\begin{small}
$\begin{array}{@{}l}
\mbox{{\bf{\tt{*PARAMETER TABLE, TYPE="UNIVERSAL\_TAB"}}}} \\
\begin{array}{ttttttt}
     2,&1,&1,&2,&1.0,&1.0,& \mu/2 \\
     4,&2,&2,&2,&1.0,&b_f,& a_f/2b_f \\
     14,&2,&2,&2,&1.0,&b_n,& a_n/2b_n \\
     6,&1,&2,&2,&1.0,&b_{fs},& a_{fs}/2b_{fs}
\end{array}
\end{array}$
\end{small}
\end{flushleft}
\paragraph{Simulation}
We incorporate all three constitutive models for myocardial tissue in the finite element analysis software solver Abaqus \cite{Abaqus2024} using our universal material subroutine, and predict the stress state of the left and right ventricular wall during diastolic filling. 
We create a finite element model of the left and right ventricular myocardial wall from high-resolution magnetic resonance images of a healthy 44-year-old Caucasian male with a height of 178\,cm and weight of 70\,kg \cite{Peirlinck2018,Peirlinck2021}. 
We spatially discretize our computational domain using 99,286 quadratic tetrahedral elements and 154,166 nodes, leading to a total 462,498 degrees of freedom. 
We compute the helically wrapped myofibers by solving a Laplace-Dirichlet problem across our computational domain, and assume a transmural fiber variation from +60$^{\circ}$ to -60 $^{\circ}$ from the endocardial to the epicardial wall \cite{Wong2012}.
The resulting microstructural organization covers 99,286 local element-based fiber, sheet, and normal vectors, 
$\vec{f}_0$, $\vec{s}_0$, $\vec{n}_0$.
We apply homogeneous Dirichlet boundary conditions at the mitral, aortic, tricupid, and pulmonary valve annuli to fix the heart in space \cite{Peirlinck2018a}, and load it with hemodynamic Neumann boundary conditions that correspond to the endocardial blood pressure during diastolic filling.
Figure \ref{figheart} showcases the resulting deformation and stress profiles in both ventricles in response to left and right ventricular pressures of 8 mmHg and 4 mmHg.
In a row-to-row comparison of the short-axis views, 
we observe small differences between the deformation invariants and the maximum principal wall stresses of all three models, with larger values for our newly discovered model and the Guan model and smaller values for the generalized myocardial Holzapfel model.
We can explain these differences by the differing goodness of fit of the three models. Moreover, we observe that our diastolic hemodynamic loading conditions enforce deformation and stress states that surpass the homogeneous tissue testing protocols of the triaxial shear and biaxial extension training data, which creates local regions of extrapolation beyond the initial training regime \cite{Martonova2024}.
Taken together, while our discovered four-parameter model best explains the laboratory experiments of triaxial shear and biaxial extension, all three models translate well into a single universal material subroutine and predict fairly similar stress profiles across the human heart.
\section{Conclusion}
In this work, we designed a universal constitutive modeling framework to predict the mechanical behavior of soft materials across a wide range of applications. 
We set up a modular material subroutine architecture which seamlessly integrates with Abaqus, and can easily be generalized towards other non-linear finite element analysis solvers. 
Doing so, our framework mitigates the risk for human error and streamlines the integration of newly discovered material models in their simulations, thus alleviating the users to perform lengthy algebraic derivations and extensive programming. 
Furthermore, our material subroutine serves as an excellent verification tool for more expert finite element software developers aiming to debug their own soft material models and finite element analysis implementations. 
We demonstrated the versatility of the universal material subroutine through numerical simulations of various living systems including the brain, skins, arteries, valves and the human heart. 
Providing a common language and material subroutine for the computational mechanics community at large, we aspire to democratize the computational analysis of soft materials amongst a broader cohort of researchers and engineers.
With one single subroutine, everyone - and not just a small group of expert specialists - can now perform reliable engineering analysis of artificial organs, stretchable electronics, soft robotics, smart textiles, and even artificial meat. 
Fostering this inclusivity, our framework can form an invaluable tool towards continued innovation and discovery in the field of soft matter overall.
%

\section*{Acknowledgements}
We thank
Kevin Linka,
Collin Haese,
Mrudang Mathur,
Denisa Martonová,
and Jiang Yao 
for their help in automated constitutive model discovery and finite element model development.
We acknowledge support through  
the NWO Veni Talent Award 20058 to Mathias Peirlinck,
the NSF CMMI Award 2320933, and
the ERC Advanced Grant 101141626 DISCOVER to Ellen Kuhl.
\noindent
\setlength{\bibhang}{0pt}
\bibliography{arxiv_universalmat.bib}
\end{document}